\documentclass[apj,a4paper]{emulateapj}
\usepackage{amssymb}
\usepackage{multirow}
\usepackage{amsmath}
\usepackage{rotating} 
\usepackage{threeparttable} 
\usepackage{pifont}
\usepackage{mathrsfs}

\shorttitle{Classification of 2XMM sources}
\shortauthors{Lo et. al.}

\newcommand{\comments}[1]{}

\begin{document}

\title{Automatic classification of time-variable X-ray sources}

\author{Kitty K. Lo$^{1,2}$,
Sean Farrell$^{1,2}$,
Tara Murphy$^{1,2,3}$, 
B. M. Gaensler$^{1,2}$
\\
$^1$ Sydney Institute for Astronomy, School of Physics, The University of Sydney, NSW 2006, Australia\\
$^2$ ARC Centre of Excellence for All-Sky Astrophysics (CAASTRO)\\
$^3$ School of Information Technologies, The University of Sydney, NSW 2006, Australia
}

\begin{abstract}

To maximize the discovery potential of future synoptic surveys, especially in the field of transient science, it will be necessary to use automatic classification to identify some of the astronomical sources. The data mining technique of supervised classification is suitable for this problem. Here, we present a supervised learning method to automatically classify variable X-ray sources in the second \textit{XMM-Newton} serendipitous source catalog (2XMMi-DR2).  Random Forest is our classifier of choice since it is one of the most accurate learning algorithms available. Our training set consists of 873 variable sources and their features are derived from time series, spectra, and other multi-wavelength contextual information. The 10-fold cross validation accuracy of the training data is ${\sim}$97\% on a seven-class data set. We applied the trained classification model to 411 unknown variable 2XMM sources to produce a probabilistically classified catalog. Using the classification margin and the Random Forest derived outlier measure, we identified 12 anomalous sources, of which, 2XMM J180658.7$-$500250 appears to be the most unusual source in the sample. Its X-ray spectra is suggestive of a ULX but its variability makes it highly unusual. Machine-learned classification and anomaly detection will facilitate scientific discoveries in the era of all-sky surveys.   

\end{abstract}


\section{Introduction}

The identification of variable and transient astrophysical sources will be a major challenge in the near future across all wavelengths. The advent of facilities such as the Large Synoptic Survey Telescope (LSST) in optical \citep{Tyson2002}, the Square Kilometre Array (SKA) in radio \citep{cordes2004} and the extended ROentgen Survey with an Imaging Telescope Array (e-ROSITA) in X-rays \citep{merloni2012}, will enable the next generation of all-sky time-domain surveys. Many types of transients and variable sources are currently known, such as supernovae, cataclysmic variables (CVs), X-ray binaries (XRBs), flare stars, gamma-ray bursts (GRB), tidal disruption flares, and future time-domain surveys will likely uncover novel source types. The large number of sources to be surveyed makes identifying interesting transients a challenging task, especially since timely multi-wavelength follow-ups will be critical for fulfilling the transient science goals. To this end, we envision that automatic classification will be a crucial part of the processing pipeline \citep{murphy2012}. 

Here, we demonstrate the feasibility of using time series and contextual information to automatically classify variable and transient sources.  We used data from the X-ray Multi-Mirror Mission - Newton (\textit{XMM-Newton})  because there has not been previous studies on this data set using automatic classification algorithms and because the time series for many of the sources are readily available, thereby enabling us to investigate the efficacy of a classifier built using solely time-domain information. Automatic classification is a similar problem across all wavelengths and we expect that the techniques used in this paper can be readily adapted for data sets in other wave-bands. 

The Second XMM-Newton Serendipitous Source Catalog Data Release 2 (2XMMi-DR2) was the largest catalog of X-ray sources \citep{watson2009} at the time it was released, but has since been surpassed by 2XMMi-DR3 and 3XMM. In this study, we used 2XMMi-DR2 and kept DR3 as a verification sample. There have been previous attempts to classify the unidentified sources in 2XMMi \citep{pineau2011, lin2012}. The traditional method is to cross-match the unknown sources with catalogs in other wavelengths (e.g. SDSS, 2MASS) and then use expert knowledge to draw up classification rules. For example, one powerful discriminant is the ratio of the optical to X-ray flux for separating active galactic nuclei (AGN) and stars. In the scheme used by \citet{lin2012}, sources whose positions coincide with the centres of galaxies are deemed to be AGN. Manually selected classification rules often have their basis in science and are usually comprehensible to other experts. This method works well when there are only a few pieces of information to be processed (e.g. optical to X-ray flux), but becomes intractable when there are many disparate sets of information. In machine learning, each piece of information is translated into either a real number or a categorical label known as a \textit{feature}. Machine learned classification excels at finding subtle patterns in data sets with a large number of features. 

Machine learned classification has been used extensively in astronomy. In X-ray astronomy, \citet{mcglynn2004} used oblique decision trees to produce a catalog of probabilistically classified X-ray sources from \textit{ROSAT}. Since that study, there have been many advances in automatic classification techniques. Ensemble algorithms such as Random Forest (RF)  have replaced single decision trees as the state-of-the-art. RF has been successfully used in astronomy for the automatic classification of variable stars \citep{richards2011, dubath2011} and the photometric classification of supernovae \citep{carliles2010}. In optical astronomy, there are efforts to incorporate automatic classification in the processing pipelines of current and planned surveys \citep{saglia2012, bloom2012, djorgovski2012}.

Feature representation is an important issue in light curve classification. Since light curves are rarely observed with exactly the same cadences, they need to be transformed into structured feature sets before different sources can be compared. Various light curve feature representations have been used in astronomy. For example, \citet{matijevi2012} transformed the light curves of each Kepler eclipsing binary into a set of 1000 observations by fitting and then interpolating the observations. However, this method only works for a very homogenous set of light curves. Other studies use a restrictive set of variability measures. In \citet{hofmann2013}, X-ray sources in M31 are placed into two light curve classes - highly variable or outbursts. This method has limited descriptive power for the variety of time-variability behaviours. In contrast, \citet{rimoldini2012} extracted a large number of features from each light curve in the Hipparcos catalog and used RF and Bayesian networks to automatically classify ${\sim}$6000 unsolved optical variables. They achieved a misclassification rate of less than 12\% and this is the methodology for feature representation that we have used.

In this paper, we present the results of using the RF algorithm to classify variable sources in 2XMMi-DR2. In Section \ref{s_data}, we describe the 2XMMi-DR2 data set and the data processing we performed. In Section \ref{s_method}, we describe the RF algorithm. In Section \ref{s_timeseries} we present the classification results using only time-series features and in Section \ref{s_contextual}, we show how the classification accuracy increases with the inclusion of contextual features. Our main result, a table of probabilistically classified 2XMMi variable sources, is presented in Section \ref{s_unknown}. In Section \ref{s_interesting} we present a method for selecting anomalous sources and briefly describe one of the interesting anomalous source. Finally, in Section \ref{s_conclusion} we discuss the limitations and future prospects of machine learned classification.  

\section{The 2XMM Variable sources}\label{s_data} 

The 2XMMi-DR2 catalog consists of observations made with the \textit{XMM-Newton} satellite between 2000 and 2008 and covers a sky area of about 420 deg$^2$. The observations were made using the European Photon Imaging Camera (EPIC) that consists of three CCD cameras -- pn, MOS1 and MOS2 -- and covers the energy range from 0.2\,keV to 12\,keV. There are $221\,012$ unique sources in 2XMM-DR2, of which $2\,267$ were flagged as variable by the XMM processing pipeline \citep{watson2009}. The variability test used by the pipeline is a $\chi^2$ test against the null hypothesis that the source flux is constant, with the probability threshold set at $10^{-5}$. 

\subsection{Data processing} 

\begin{figure} 
\begin{center} 
\includegraphics[width=0.45\textwidth]{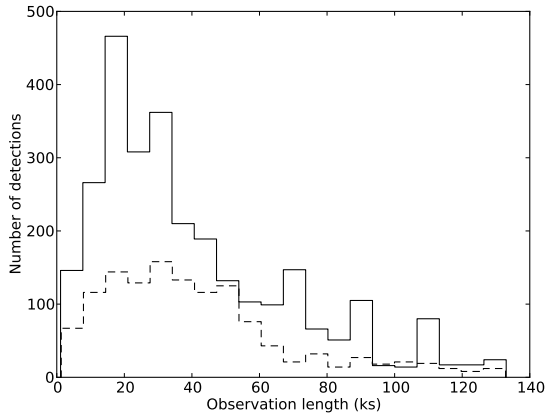} 
\caption{Number of detections vs. observation lengths for the light curves in our sample. Solid line is the classified sample in the training set; dotted line is the unidentified sample in the test set.} 
\label{fig:obs_length}
\end{center} 
\end{figure}

In this paper, a detection refers to a light curve in an epoch made by one camera. Each detection in our sample has an associated light curve which consists of background subtracted count rates, count rate errors, background count rates, background errors, and time stamps. A source can be detected in multiple epochs, and in each epoch there are typically three detections, one by each of the pn, MOS1 and MOS2 cameras. The exposure time per detection ranges from a few ks to over 100~ks (Figure \ref{fig:obs_length}). The bin widths are in multiples of 10\,s and are large enough such that there are a minimum of 18 counts/bin and 5 counts/bin for the pn and MOS detectors respectively. To ensure that all the variability in the light curve comes from the source and is not due to background flares or instrumental errors, we filtered out points likely to contain errors. First, we removed all points that lie outside the good time intervals (GTIs). GTIs are time periods where monitored parameters, such as spacecraft attitude stability and background particle levels, are within acceptable levels. Second, we removed all points where the fraction of time exposed, $F_{exp}$, was $<0.9$. Count rates determined during a low $F_{exp}$ measurement are not reliable. Third, we removed points with zero error rates. Since an error of zero is not realistic, it indicates some error in the data processing or the observation. After the filtering step, we removed sources from the sample with less than 15 data points in the light curve. Table \ref{tab:sample} is a breakdown of the sources in our sample. In total, we excluded 983 sources from further considerations.    

\begin{table}
\caption{The variable 2XMMi sample} 
\begin{center} 
\begin{tabular}{lr} 
&  Sources \\
\hline 
\hline 
Total excluded from our sample & 983\\
\hspace{0.5cm}Spurious & 924 \\ 
\hspace{0.5cm}Classified - not enough data points & 14 \\ 
\hspace{0.5cm}Classified - classes with few sources & 17 \\ 
\hspace{0.5cm}Unidentified - not enough data points & 28 \\
\hline
& \\
Total in our sample & 1284 \\
\hspace{0.5cm}Classified - in the training set &  873 \\
\hspace{0.5cm}Unidentified - in the test set & 411 \\ 
\hline
Total variable sources & 2267 \\
\hline \\
\end{tabular}
\end{center} 
\label{tab:sample}
\end{table} 

\subsection{Classified sample} 

For our training set we used the classifications for each discrete variable 2XMMi-DR2 source as determined by Farrell et al. (in prep). While the classification methodology will be discussed in detail in Farrell et. al. (in prep), we summarise the process as follows. First, the pipeline produced images, spectra, and light curves were manually inspected using the products available on the Leicester Database and Archive Service (LEDAS) webpages\footnote{\url{http://www.ledas.ac.uk/}}. Spurious detections were identified, primarily  through examination of the images, and summarily discarded. Detections of extended sources were also discarded (e.g. supernova remnants, galaxy clusters etc.) as any variability detected from these sources within a single XMM-Newton observation would have to be spurious. In this manner we discarded 924 out of the original 2,267 variable sources as spurious.

The nature of the remaining 1,343 real variable sources was determined by searching for matches around the source
positions in the SIMBAD astronomical database\footnote{\url{http://simbad.u-strasbg.fr/simbad/}} and the NASA/IPAC Extragalactic Database\footnote{\url{http://nedwww.ipac.caltech.edu/}} (NED), and through a shallow review of the literature. The bulk of these sources (44\%) were associated with stars, with the rest associated with the centres of galaxies (i.e. AGN; 7\%), XRBs (6\%), CVs (6\%), ultraluminous X-ray sources (ULXs; 1\%), GRBs (1\%), and super soft sources (SSSs; 1\%). A very small number (representing $\sim$ 1\%) were associated with planets (Jupiter and Saturn), extragalactic globular clusters, and magnetars. The remaining sources, comprising 33\% of the real variable source sample, did not have a match in either SIMBAD or NED and are thus unidentified. The training set thus contains 873 sources in seven classes: AGN, CVs, GRBs, XRBs, SSSs, stars, ULXs, and XRBs, with the unidentified sources not included. Table 2 shows a breakdown of the number of sources and detections we have in the classified training set and Figure 2 shows examples of light curves from each class.


AGN are the central regions of galaxies believed to contain supermassive black holes. X-ray emission from AGN is mainly due to inverse Compton scattering and typically follows a power-law spectrum \citep{longair}. We included different types of stars under the ``star'' category, including flare stars, binaries, pre-main sequence stars and young stellar objects. Late-type flare stars produce X-ray emission from magnetic reconnection in their coronae \citep{benz2010}. A CV is a binary system in which a white dwarf accretes from a companion star. The typical orbital periods of CVs are between 75~min and 8\,hrs. CVs can be magnetic (mCV) or non-magnetic; the former are also known as polars or intermediate polars. X-ray emission from non-magnetic CVs is mainly due to low temperature thermal plasma emission from shocks formed when material accretes onto the white dwarf. In mCVs, the accretion disk is suppressed by the magnetic field and the X-ray emission arises from the boundary of the shock of the collimated accretion flow. XRBs are binary systems where the accreting compact object is a black hole or neutron star. The donor star in a high-mass XRB is usually a massive O or B-type star, or a Be star while the donor star in a low-mass XRB can be a main-sequence star, a white dwarf or a red giant. Both subtypes of XRBs are included in this category. ULXs are objects with X-ray luminosities exceeding that of a stellar mass black hole accreting at the Eddington limit. They are located within galaxies but not in the nucleus regions. SSSs, as their name suggests, are characterised by their extremely soft (peaking at $<0.5$keV) spectra. The accepted paradigm for their nature is that of a white dwarf binary with steady nuclear burning \citep{kahabka2006}. Lastly, the GRBs we are referring to here are afterglow emission from long GRBs, which are believed to be the core collapses of massive stars.  

\begin{figure} 
\begin{center} 
\begin{tabular}{cc}
    \includegraphics[width=0.24\textwidth]{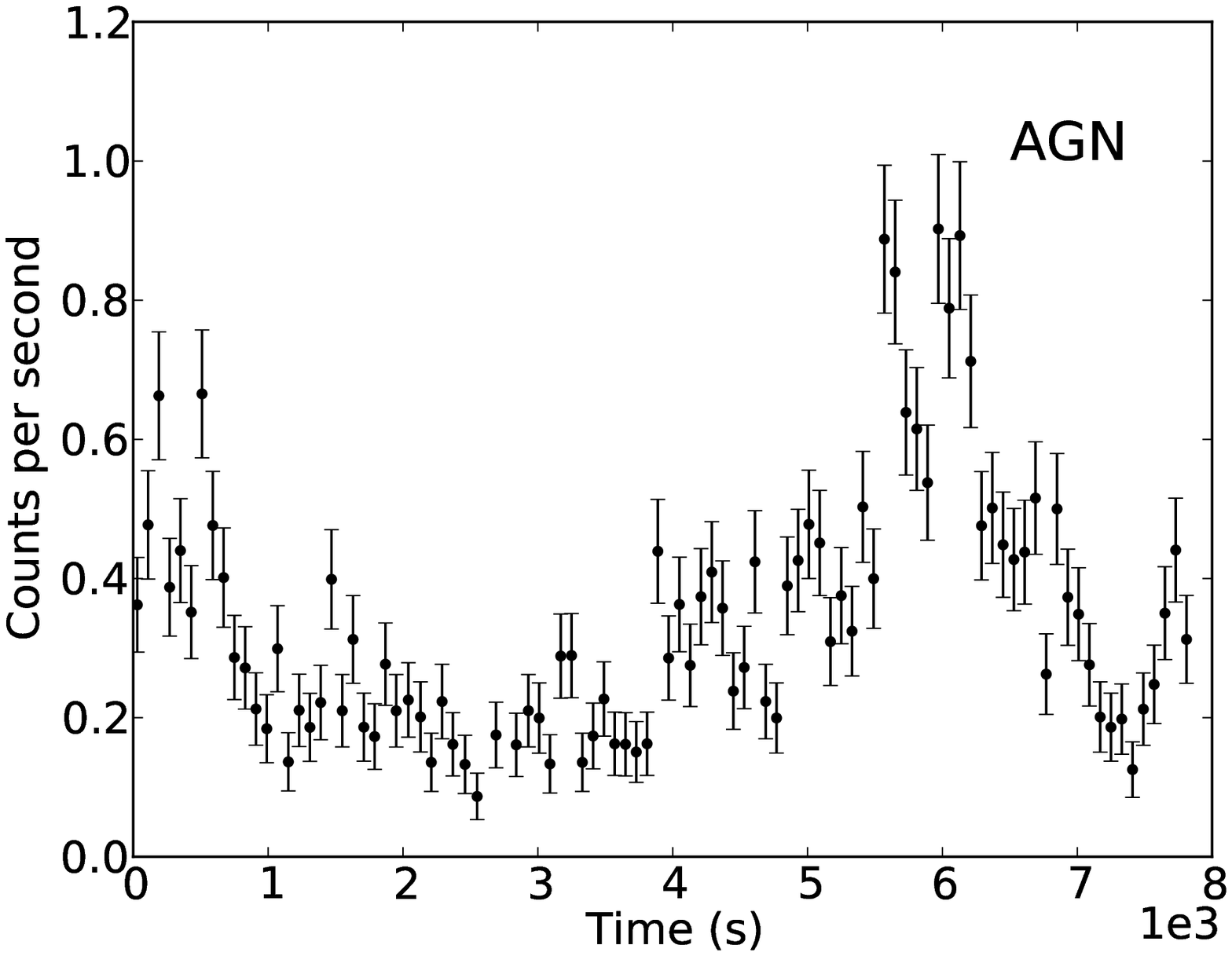}&

    \includegraphics[width=0.24\textwidth]{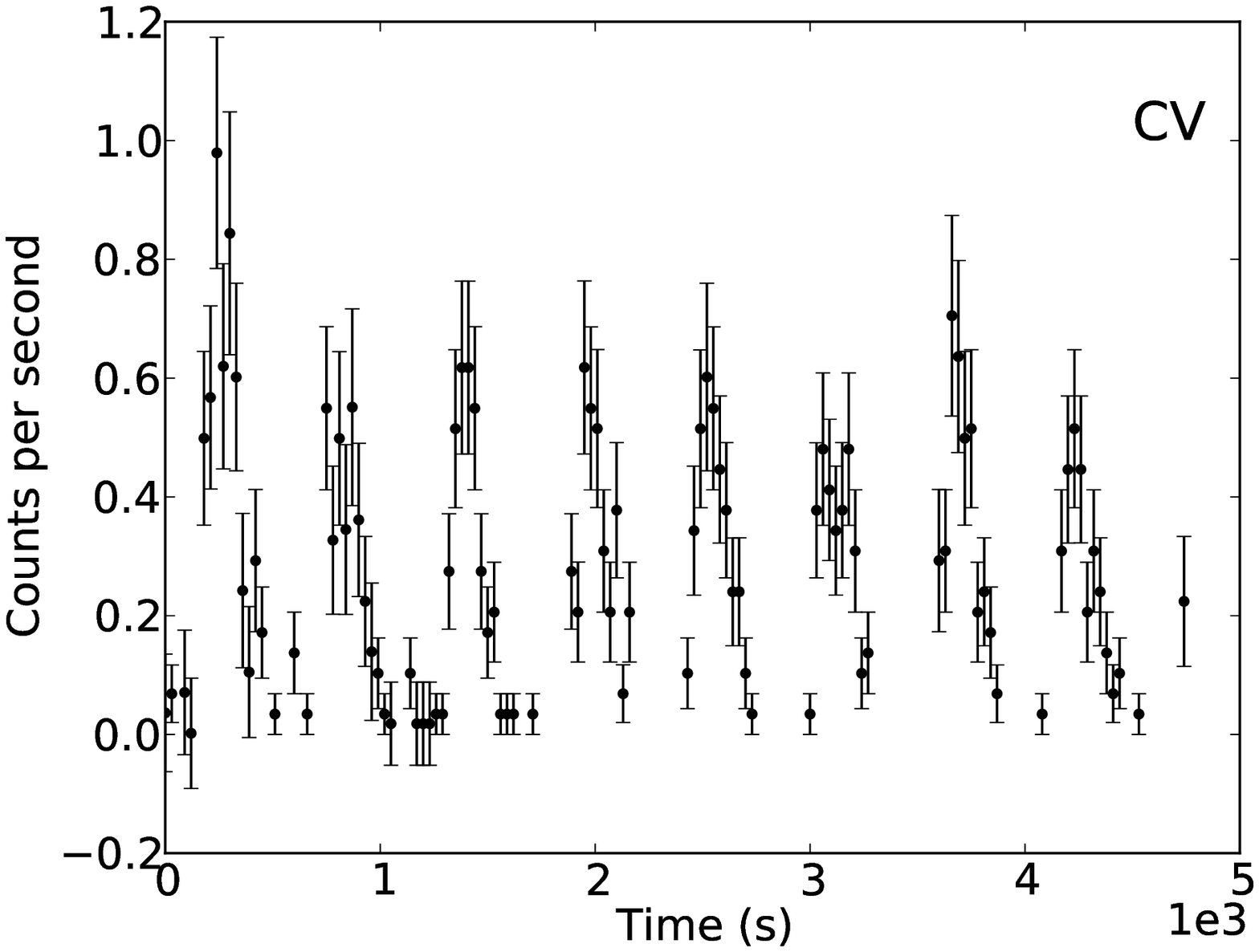}\\

    \includegraphics[width=0.24\textwidth]{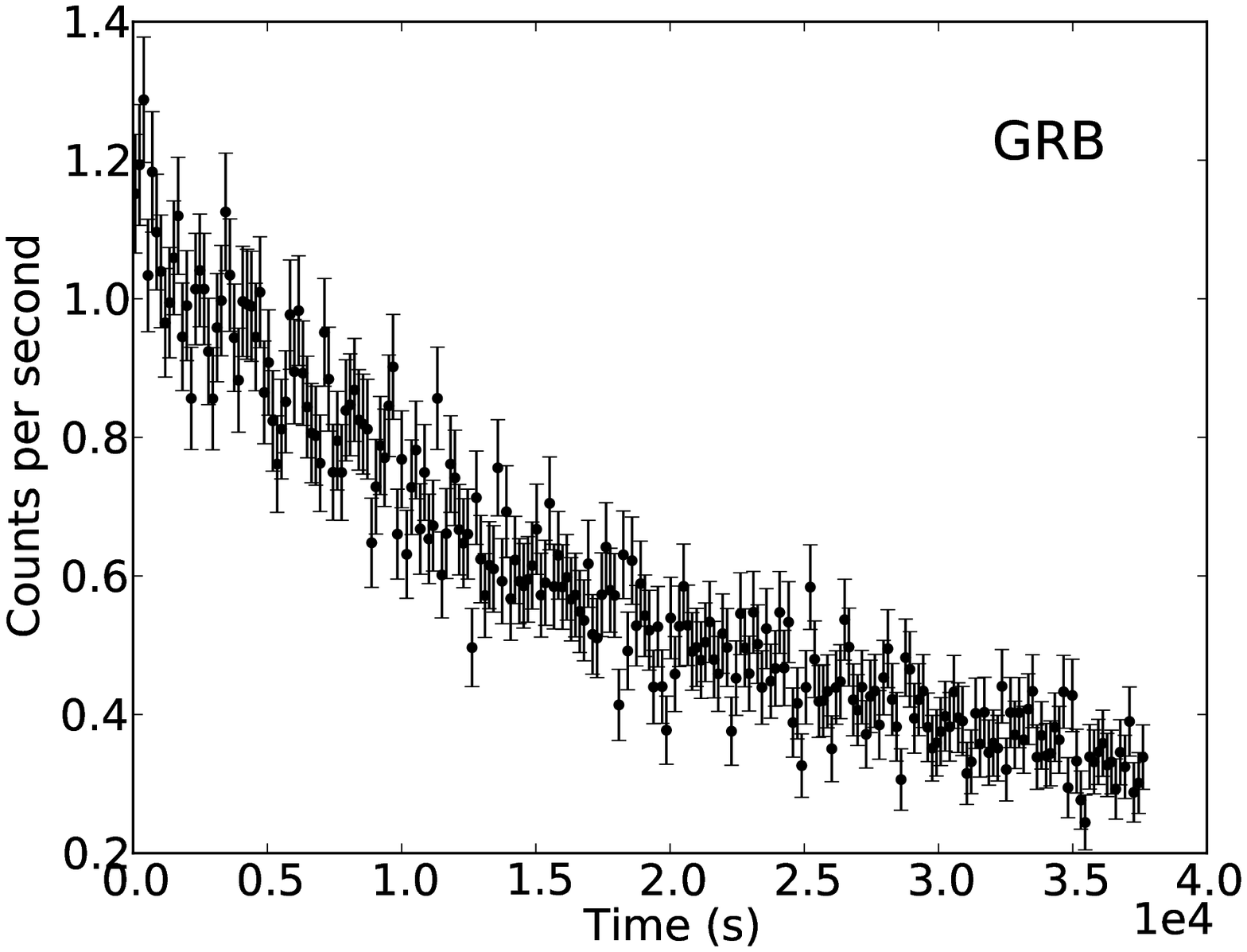}&

    \includegraphics[width=0.24\textwidth]{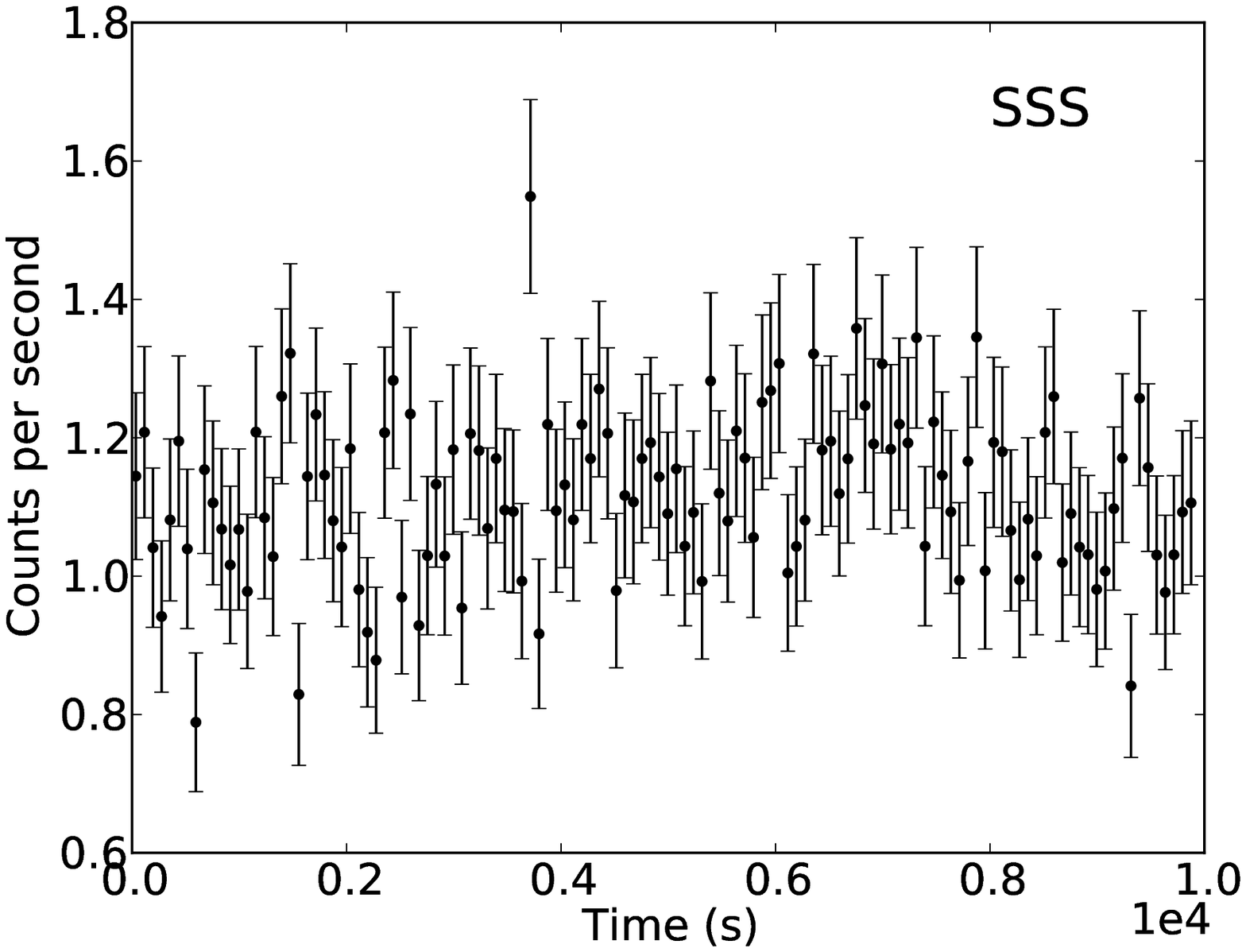}\\
                
    \includegraphics[width=0.24\textwidth]{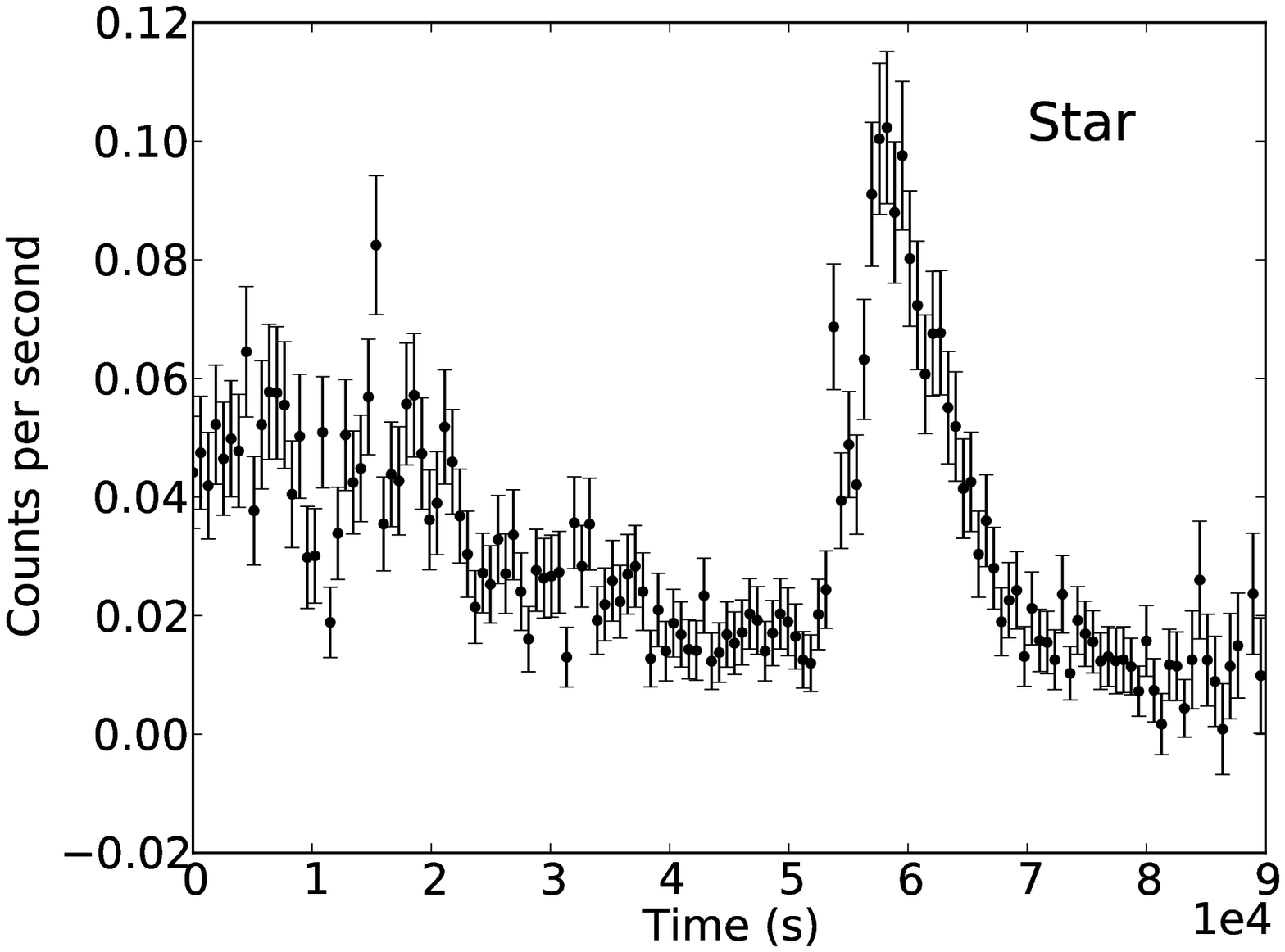}&
    
    \includegraphics[width=0.24\textwidth]{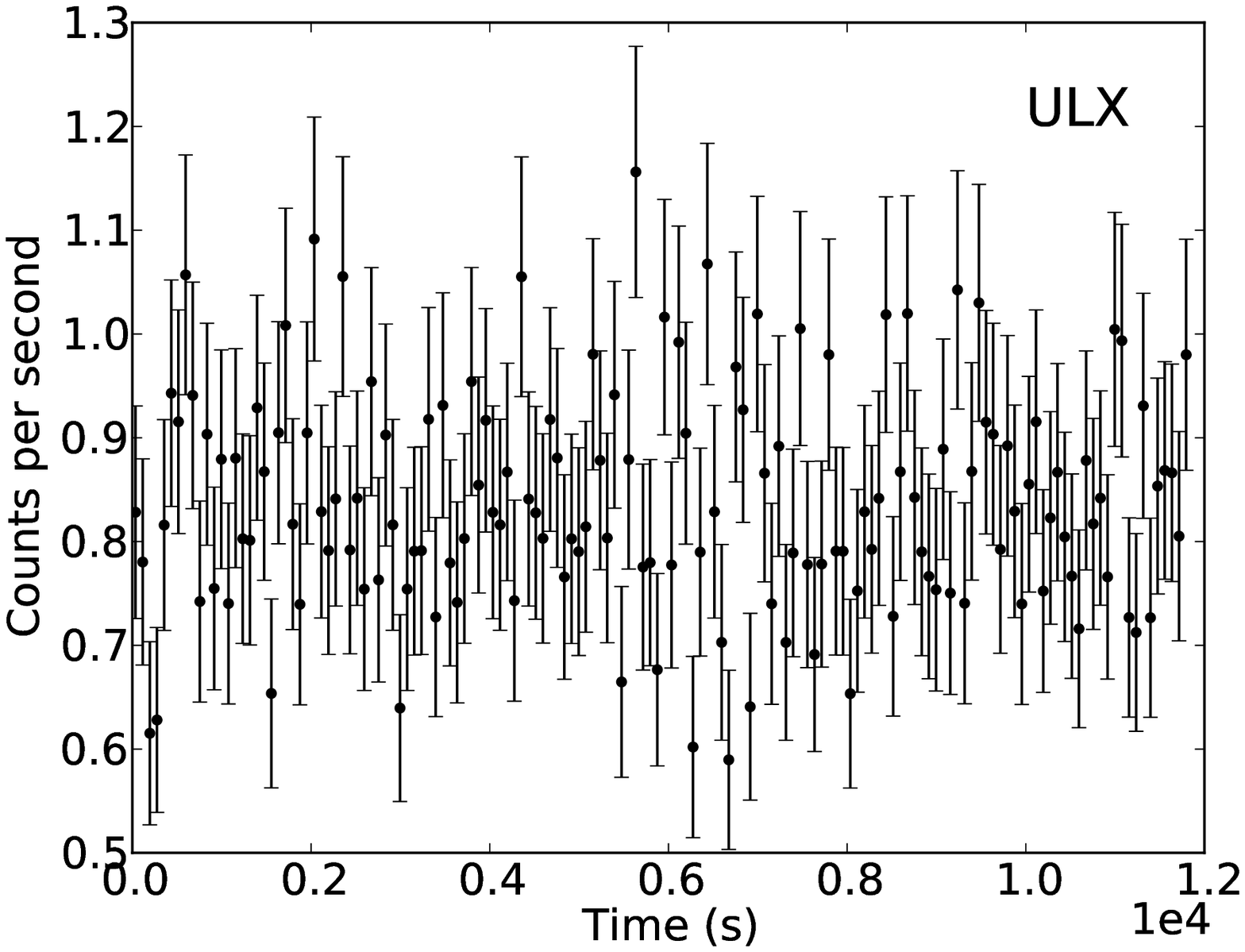}\\
        
    \includegraphics[width=0.24\textwidth]{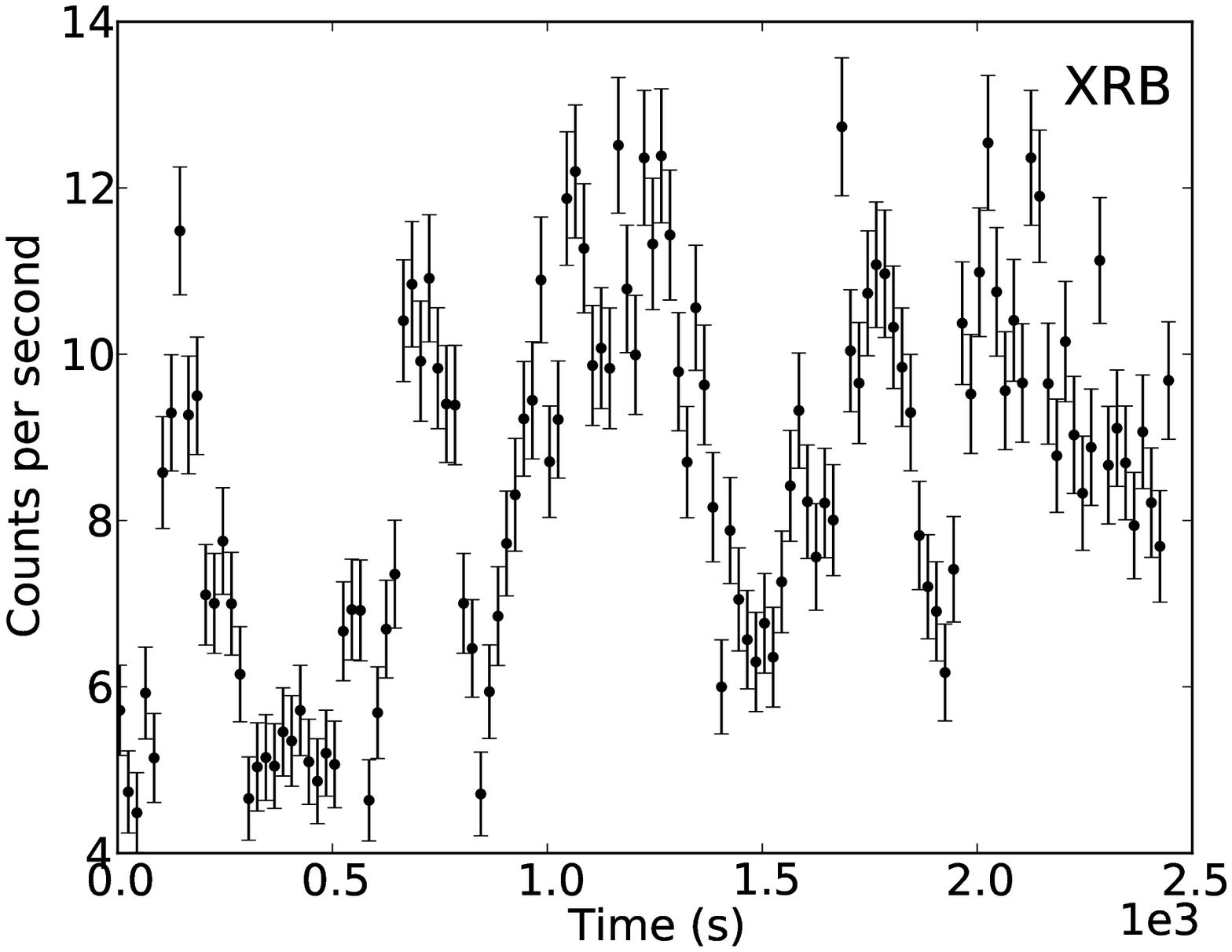}&\\
    
  \end{tabular}
\caption{Example lightcurves for the seven types of X-ray sources in our training set.}
\label{fig:lc_examples} 
\end{center} 
\end{figure} 

\begin{table}
\caption{Classified sources in the training set} 
\begin{center} 
\begin{tabular}{crr} 
Class & Detections & Sources \\
\hline 
\hline 
AGN  & 298 & 99\\
CV   & 131 & 83\\
GRB  & 9  & 9 \\
SSS  & 19  & 8\\
Star & 1953 & 573\\
ULX  & 87  & 17 \\
XRB  & 322 & 84\\
\hline 
Total & 2819 & 873 \\
\hline 
\end{tabular}
\end{center} 
\label{tab:class}
\end{table} 

\section{Classification method} \label{s_method} 

\subsection{Introduction} 

The machine learning technique we use here is known as supervised classification \citep{duda2001}. Supervised classification uses a set of labelled training examples to construct a prediction model. In Section \ref{s_data}, we described the method used to construct the training set. In this section we will explain the classification algorithm in detail. 

\subsection{Random Forest} 

Random Forest (RF) is an ensemble supervised classification algorithm developed by \citet{breiman2001}. 
In the training phase, the algorithm builds an ensemble of decision trees. Each tree is built using a bootstrap sample from the training set, i.e. for $S$ samples in the training set, the algorithm randomly picks $S$ samples with replacement to create the training set for each tree. To construct a decision tree, training samples are split at each \textit{node} (a node is where classification decisions are made) and this process iterates until all the training samples at the node belong to the same class. The feature used at each node is the one that produces the highest decrease in Gini impurity, as calculated using the equation: 

\begin{equation}
\label{eq:gini} 
G = \displaystyle\sum_{i=1}^{m}f_k (1-f_k) , 
\end{equation} 

where $m$ is the number of classes and $f_k$ is the fraction of sources which belongs to class $k$. The Gini impurity becomes zero when all the sources in a node are of the same type. In RF, a small subset of features (typically only a small fraction of the total number of features) are randomly chosen to be considered at each node. To predict the class of a new sample, each decision tree in the ensemble votes for a class and the output class is the one with the most votes. 

RF is one of the most accurate classification algorithms available \citep{caruana2006}. It can handle large datasets with large number of features. RF can generalize without overfitting; an overfitted model is one that describes noise rather than the true underlying relationship between features. It is also simple to optimize, since there are only two parameters to adjust -- the number of trees and the number of variables to use at each node. We used the \texttt{R} package \citep{rpackage} \texttt{randomForest} \citep{liaw2002} for the experiments performed in this paper. Using tuning function \texttt{tuneRF} in the \texttt{randomForest} package, we found that the optimal number of variables to use at each node is 9. To find the optimal number of trees, we repeated the experiment with different number of trees, and found 500 trees was optimal. 

\subsection{Unbalanced training set} 

Our training set, as summarized in Table \ref{tab:class}, is heavily unbalanced. Stars, the most abundant class, outnumber GRB, the rarest class by around 200 to 1. Heavily unbalanced training sets can degrade the performance of a machine learned classification algorithm. To ameliorate this issue, we oversampled the two most under-represented classes - GRB and SSS, using the SMOTE algorithm \citep{chawla2002}. SMOTE creates synthetic minority class samples by using the k-nearest neighbours and has been shown to be more robust than simply oversampling the minority class with replacement. We used the SMOTE implementation in the \texttt{DMwR} package \citep{dmwr} in \texttt{R} to oversample the GRB class by ten fold and the SSS class by four fold. 

\subsection{Class membership probabilities} 

Class membership probabilities can be more informative then discrete class labels. The former provides information on the degree of confidence of the classification, and allows the user to set cutoffs for selecting their class of interest based on the desired level of reliability and completeness. RF can provide class membership probabilities in the form of the fraction of votes in the ensemble given for the class. In this paper, we report all results as class membership probabilities. 

\section{Classification with time series features}\label{s_timeseries}

The variable X-ray source sample allows us to investigate the usefulness of variability information in classification. In this section, we describe the time series features we extracted from the X-ray light curves and report on the accuracy of the RF classifier trained using only time series features. Table \ref{tab:char} is a summary of the general light curve characteristics of each source type. Although we cannot arrive at a definitive classification solely using the light curves, variability information can narrow down the potential classes. For example, a source with periodic variability is highly unlikely to be an AGN, but it could be a CV or an XRB. 

Based on the expected variability characteristics, we extracted four types of light curve features -- periodic features, likelihood of power law decay, flares and statistical features such as fractional variability. Table \ref{tab:ts_features} is a summary of the time series features. We discuss each feature in detail in the following sections. 

\begin{table}
\caption{Light curve characteristics of each source type} 
\begin{center} 
\begin{tabular}{cp{0.3\textwidth}} 
Class & Light curve characteristics \\
\hline 
\hline 
AGN  & Flickering, stochastic aperiodic variability\\
CV   & Some sources display periodicity\\
GRB  & Power law decay\\
SSS  & Typically constant, occasional variability\\
Star & Flares and bursts lastings minutes\\
ULX  & Typically constant, occasional variability\\
XRB  & Some sources display periodicity, flickers and flares\\
\hline
\end{tabular}
\end{center} 
\label{tab:char}
\end{table} 

\begin{table*}
\caption{List of time series features used for classification} 
\label{tab:ts_features}
\begin{center} 
\begin{tabular}{lp{12cm}}
\hline
Feature & Description \\
\hline 
\hline 
Lomb-Scargle\_amp1 & Amplitude of the best-fitting sine function at the highest Lomb-Scargle periodogram peak\\
Lomb-Scargle\_amp2 & Amplitude of the best-fitting sine function at the second highest Lomb-Scargle periodogram peak\\
Lomb-Scargle\_period1 & Period in seconds corresponding to the highest Lomb-Scargle periodogram peak \\
Lomb-Scargle\_period2 & Period in seconds corresponding to the second highest Lomb-Scargle periodogram peak \\
Lomb-Scargle\_FAP1 & False alarm probability of the highest Lomb-Scargle periodogram peak \\
Lomb-Scargle\_FAP2 & False alarm probability of the second highest Lomb-Scargle periodogram peak \\
Powerlaw\_C & Parameter C in the best fit power-law model $y(t)=F_0 \left(t-t_0\right)^{-C}$\\
Powerlaw goodness of fit & Reduced $\chi^2$ statistics of the exponential model \\ 
Flare\_nums & Number of flares found \\
Flare\_amp & Amplitude of the strongest flare \\
Flare\_duration & Duration in seconds of the strongest flare \\
Amplitude & 0.5 $\times$ [Max(count) - Min(count)]\\
Standard deviation\_dev & Standard deviation of the counts\\
Beyond1Std  & Percentage of observations that lie beyond one standard deviation from the weighted mean \\
Flux ratio mid 20\% & Ratio of flux in the 60th to 40th percentiles over 95th to 5th percentiles\\
Flux ratio mid 35\% & Ratio of flux in the 67.5th to 32.5th percentiles over 95th to 5th percentiles\\
Flux ratio mid 50\% & Ratio of flux in the 75th to 25th percentiles over 95th to 5th percentiles\\
Flux ratio mid 65\% & Ratio of flux in the 82.5th to 17.5th percentiles over 95th to 5th percentiles\\
Flux ratio mid 80\% & Ratio of flux in the 90th to 10th percentiles over 95th to 5th percentiles\\
Skew & Skew of the distribution of count rates; calculated using the python function scipy.stats.skew \\
Max slope & Maximum slope of adjacent observation points [counts s$^{-1}$] \\
Med\_abs\_dev & Median of the absolute value of the deviation from the median \\
Med\_buffer\_range\_percentage & Percentage of measurements within 20\% of the median \\
Percentage\_amp\_diff & Maximum difference between a measurement and the median as a percentage of the median\\
Percentile\_diff & Count rate at the 98th percentile minus the count rate at the 2nd percentile \\
Modulation Index & variance / weighted mean \\
Fractional var & Fractional rms variability, calculated as $\frac{1}{\bar{y}}\sqrt{\frac{\sum \left(y_i - \bar{y} \right)^2 - \sum \sigma_{i}^2}{N}}$, where $y_i$ is the count rate at time $i$, $\sigma_{i}$ is the error of the count rate at time $i$, $\bar{y}$ is the average count rate and $N$ is the number of observations in the time series.\\ 
\hline 
\end{tabular}
\end{center} 
\end{table*}

\subsection{Periodic features} 

Some CVs and XRBs display periodicities on timescales of minutes to hours, less than the typical length of our observations \citep{israel2002, hearn1977}. This suggests the frequency domain can inform our classification. We used the generalised Lomb-Scargle periodogram from \citet{zechmeister2009} to represent the frequency domain information. The advantage of this technique over a conventional Fourier transform is that it can handle unevenly sampled data. For evenly sampled light curves, this would be unnecessary. However, due to the filtering process, our light curves may be missing data points. The generalised periodogram is equivalent to fitting functions of the form $y = a\cos{\omega t} + b\sin{\omega t} + c$. The inclusion of the offset $c$ makes it more general than the original Lomb-Scargle periodogram \citep{lomb1976}. Finding the best fit translates to minimizing the squared difference between the data at time $i$, $y_i$, and the model $y(t_i)$ represented by the $\chi^2$ function: 

\begin{equation}
\label{eq:chisq} 
\chi^2 = \sum \frac{\left[ y_i - y(t_i) \right]^2}{\sigma_i^2},
\end{equation} 

where $\sigma_i^2$ is the estimated variance at time $i$. 

The periodogram can be written as: 
\begin{equation}
\label{eq:LSP} 
P(\omega) = \frac{N-1}{2}\frac{\chi_0^2 - \chi^2(\omega)}{\chi_0^2}, 
\end{equation} 

where $\chi_0$ is the squared deviation of $y_i$ from the mean. 

Equation (\ref{eq:LSP}) has been normalized by the factor $\left(N-1\right)/2$ ($N$ is the number of measurements in the time series) so that if the data are pure noise, then the expected periodogram value is 1. This equation has an analytical solution \citep[Equation (5) in ][]{zechmeister2009} that we used to calculate the periodogram value. The false alarm probability (FAP) is also included in our feature set as a way to capture the significance of the periodogram value. FAP is calculated using: 

\begin{equation}
\label{eq:fap} 
\textrm{FAP} = 1 - \left[ 1 - \left(1-\frac{2P}{N-1}\right)^{\frac{N-3}{2}}\right]^M , 
\end{equation}

where $M$ is the number of peaks in the periodogram. This relies on an implicit assumption that the noise in the flux is Gaussian. 

\begin{figure} 
\begin{center} 
\includegraphics[width=0.5\textwidth]{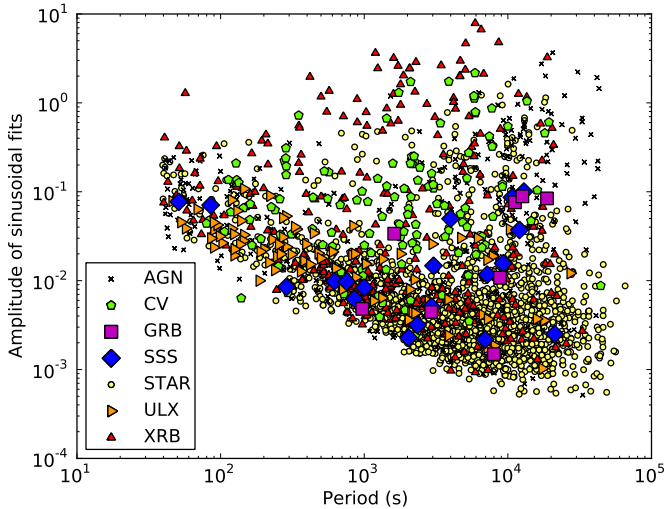} 
\caption{Amplitude of the best fitting sine function from the Lomb-Scargle periodogram for the training set plotted against the period in seconds.} 
\label{fig:period}
\end{center} 
\end{figure} 

For our classification experiments, we only used the two highest peaks in the periodogram. For each peak, we extracted the amplitude of the best fitting sine function ($\sqrt{a^2+b^2}$), the period in seconds and the FAP. Figure \ref{fig:period} shows a plot of the first two of these three values. To ensure the second peak is truly distinct from the first, we eliminated values immediately adjacent to the highest peak and found the second highest peak from the remaining frequency bins. 

\subsection{Fit to a power-law model} 

\begin{figure} 
\begin{center} 
\includegraphics[width=0.5\textwidth]{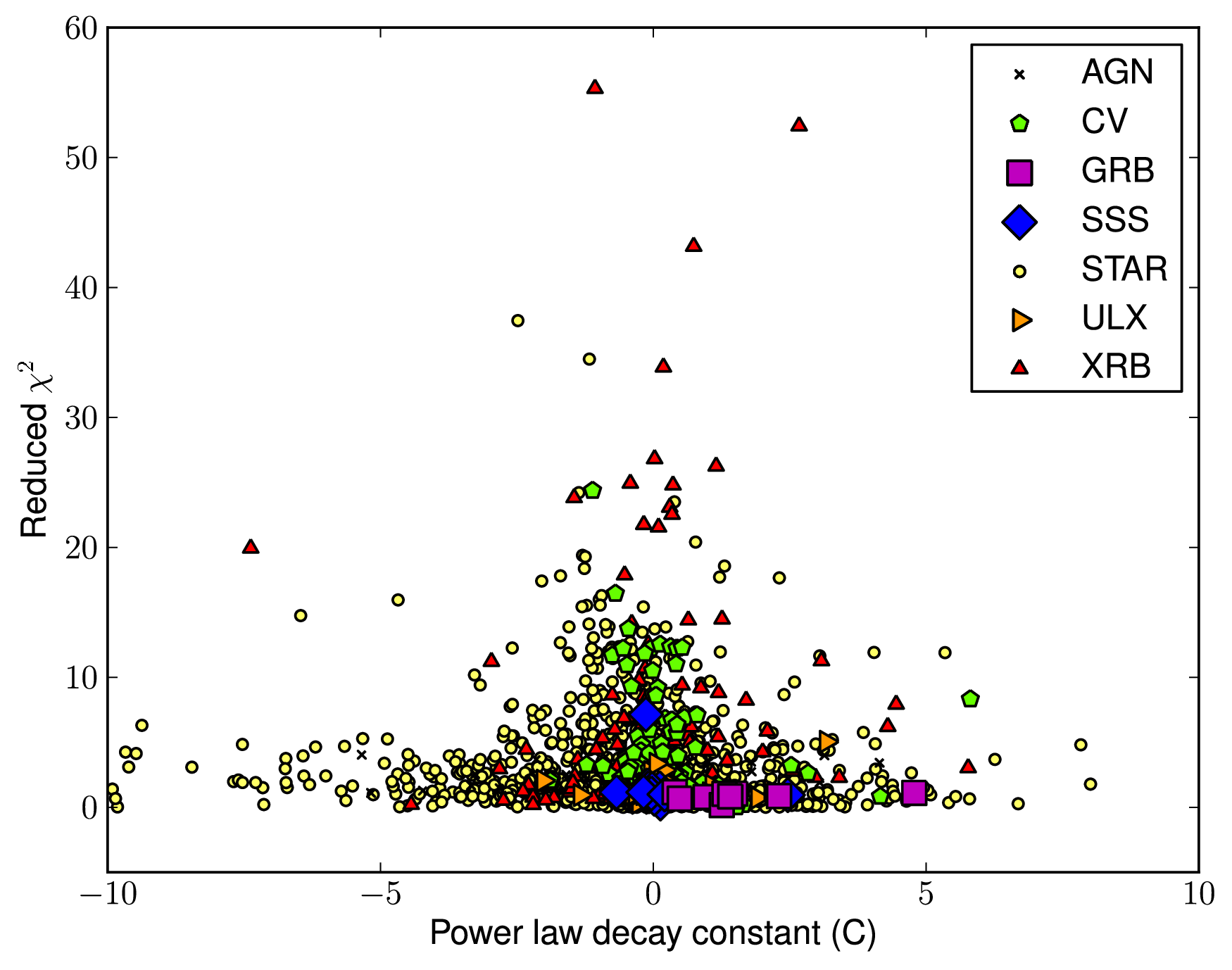} 
\caption{Plot of the reduced $\chi^2$ from an power law fit to the light curve, and the decay constant of the fit.} 
\label{fig:exponential}
\end{center} 
\end{figure}

The identifying feature of a GRB afterglow light curve is a power-law function. We fitted a power-law function of the form $y(t)=F_0 \left(t-t_0\right)^{-C}$ to the light curves and used the parameters $C$ of the best fitting model as classification features. The fitting procedure also determined $F_0$ and $t_0$ but these were not used as classification features. We used the \texttt{curvefit} function from the Python package \texttt{scipy} to perform the least squares non-linear fit. This process assumes the input errors are Gaussian, which is not always satisfied due to low count rates. To circumvent this issue, we binned the data to coarser time bins such that the average number of counts per bin was at least 20. To estimate the goodness of fit, we calculated the $\chi^2$ statistic using $\left(y_i-\hat{y}\right)^2 / \sigma_i^2$, where $\hat{y}$ is the model estimate of $y_i$ and $\sigma_i$ is the error after binning. The reduced $\chi^2$ is another feature for our classifier (Figure \ref{fig:exponential}). 

\subsection{Flare finding} 

X-ray flares are common features in active stars. To test for the existence of flares, we decomposed each light curve into a piecewise constant representation and then looked for segments with elevated count rates compared to adjacent segments. We used the Bayesian blocks technique to construct the piecewise constant segments \citep{scargle1998}. This technique is designed for astronomical count data with Poissonian noise and is based on the Bayesian formalism. It relies on comparing two hypotheses -- the unsegmented hypothesis where the light curve can be described with one rate, and the segmented hypothesis where the light curve is described with two rates. The likelihood that the count rate is constant is given by: 

\begin{equation}
L\left(H_{unseg} | Data\right) = \frac{\Gamma\left(A+1\right)}{\left(B+1\right)^{A+1}},
\end{equation}

from Equation (29) in \citet{scargle1998}, where $A$ is the number of photons and $B$ is the number of bins. On the other hand, the likelihood of the segmented model is:

\begin{equation}
L\left(H_{seg} | Data\right) = \frac{\Gamma\left(A_1+1\right)}{\left(B_1+1\right)^{A_1+1}} \times \frac{\Gamma\left(A_2+1\right)}{\left(B_2+1\right)^{A_2+1}},
\end{equation} 

where $A_1$, $B_1$ and $A_2$, $B_2$ are the number of photons and number of bins in segment one and segment two respectively. To compare the two hypotheses, we calculated the odds ratio: 

\begin{equation}
O_{12} = \frac{L\left(H_{unseg} | Data\right)}{L\left(H_{seg} | Data\right)}. 
\end{equation}

\begin{figure} 
\begin{center} 
\includegraphics[width=0.5\textwidth]{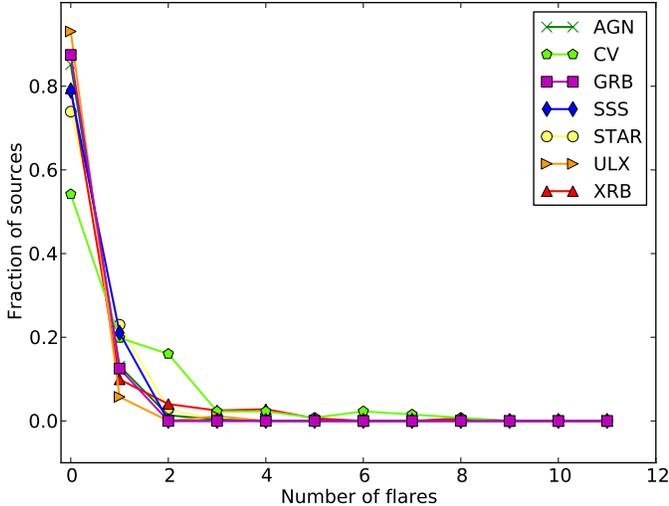} 
\caption{Fraction of sources of each type in the training set according to the number of flares found.} 
\label{fig:flares}
\end{center} 
\end{figure}

If $O_{12}$ is less than one, then a segmented model is favored and the change point that yields the highest likelihood is used to segment the light curve. This process is performed recursively and terminates when further segmentation no longer improves the likelihood.  We then count the number of segments where the count rate is at least three times higher than the expected error compared to the preceding and succeeding segments. This is our count of the number of flares in the light curve as shown in Figure \ref{fig:flares}. 

\subsection{Statistical features} 

In addition to the light curve features described in the previous sections, we extracted 16 statistical features that were used by \citet{richards2011} in the classification of variable stars. These are general statistical measures that do not depend on the time ordering of the measurements, e.g. fractional variability, mean, and standard deviation. Detailed descriptions of these features can be found in Table \ref{tab:ts_features}.  

\subsection{Accuracy on training set} 
\label{sec:acc_timeseries} 

To evaluate the accuracy of our classifier, we used the method of cross-fold validation. We divided the training sample into ten sets, trained with nine sets, used the model created to classify the remaining sample set and then repeated for ten different combinations. The overall accuracy is the total number of correctly classified samples divided by the total number of samples in the training set. Using only time-series features, the overall accuracy is $\sim77\%$. 

\begin{figure} 
\begin{center} 
\includegraphics[width=0.5\textwidth]{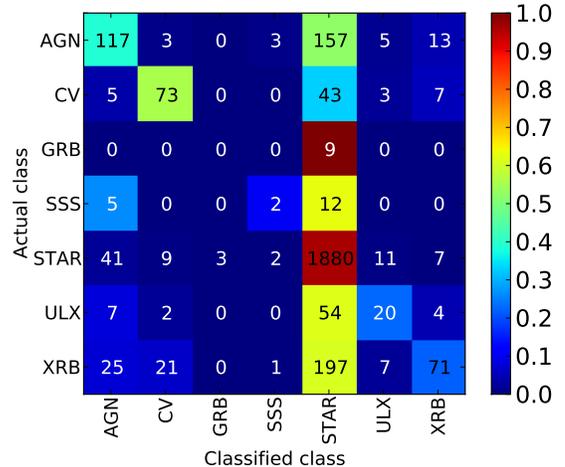} 
\caption{Confusion matrix from performing 10-fold cross-validation on the training set using the RF classifier with only time-series features. The color bar represents the true positive rate.} 
\label{fig:conf_training}
\end{center} 
\end{figure}

Figure \ref{fig:conf_training} shows the confusion matrix, where the number in each square represents how the detections are classified. The sum of each row of the confusion matrix is the total number of detections in that class. The numbers in the diagonals are detections that have been correctly classified. GRBs, SSSs, and ULXs are the three worst performing classes. This is not unexpected since SSSs and ULXs have no distinguishing time series features. In contrast, stars, XRBs, and CVs performed relatively well and are usually only confused with each other. From Figure \ref{fig:lc_examples}, it can be seen that XRBs and CVs share semi-periodic temporal behaviour while stars have distinguishing flares. It is also worth noting that sources of all types are most likely to be mis-classified as stars. Since a significant proportion of our training set are stars, the classifier optimises for accuracy by labelling sources for which it does not have sufficient information as the majority class. 

\begin{figure} 
\begin{center} 
\includegraphics[height=0.4\textwidth]{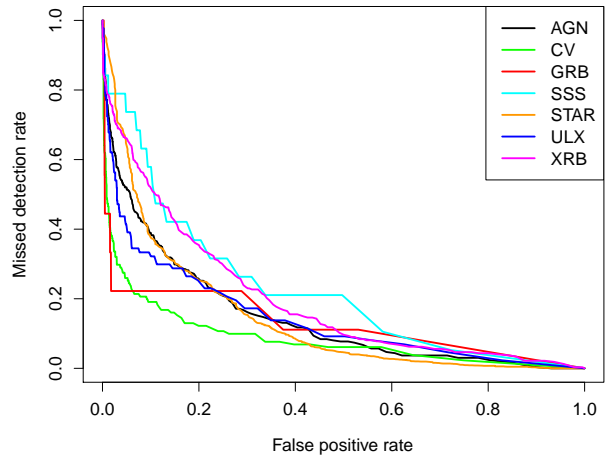} 
\caption{ROC plot from performing 10-fold cross-validation on the training set using the RF classifier with only time-series features.} 
\label{fig:roc_training}
\end{center} 
\end{figure}

Figure \ref{fig:roc_training} shows a plot of the missed detection rate vs. the false positive rate, known as the Receiver Operating Characteristic (ROC) plot. Missed detection is 1 - true positive rate, which is the proportion of samples classified as the actual class; false positive rate is the proportion of samples not of the class but classified as such. We created the ROC plot by transforming the results of the multi-class classification into a binary classification, i.e. each sample either belongs to the actual class (true positive) or it does not (false positive). A well performing classifier should be able to provide a low missed detection rate with a low false positive rate, i.e. be on the bottom-left part of the ROC plot. Figure \ref{fig:roc_training} shows CVs are the best performing source type, even though they only achieved an accuracy of 52\%. This accuracy is lower than what may be expected since the missed detection rate is only ${\sim}10\%$ when the false positive rate is ${\sim}20\%$. This is because the test set is unbalanced, a small number of stars mis-classified as CVs would not significantly decrease the accuracy for stars, and therefore would not lead to a high false positive rate.

\begin{figure} 
\begin{center} 
\includegraphics[height=0.6\textwidth]{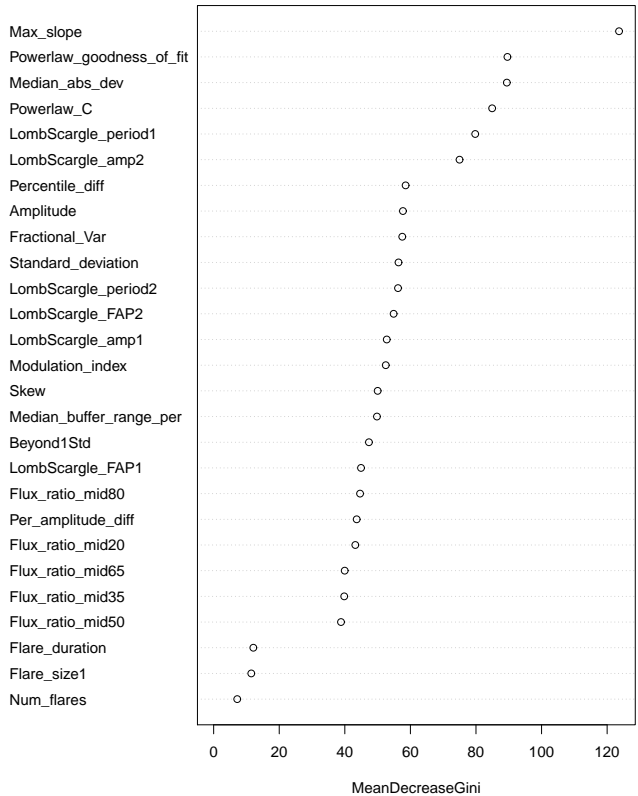} 
\caption{Relative importance of the time series features. The features are described in detail in Tables \ref{tab:ts_features}.} 
\label{fig:varImp_tonly}
\end{center} 
\end{figure}

The \texttt{R} package \texttt{randomForest} also has the ability to calculate relative feature importance. The importance of each feature is estimated by calculating the total decrease in Gini impurity (Equation \ref{eq:gini}) from using that feature, averaged over all the trees in the forest. Figure \ref{fig:varImp_tonly} shows the mean decrease in Gini impurity for the time series features. The five most relevant features are, in order of importance: max\_slope, powerlaw\_goodness\_of\_fits, median\_abs\_dev, powerlaw\_C, and LombScargle\_period1 (all as defined in Table \ref{tab:ts_features}). 

\section{Classification with contextual features} \label{s_contextual}

In Section \ref{s_timeseries}, we showed that time-series features have some discriminative power, but that the classification accuracy is insufficient for practical use. In this section, we expand our feature set to include hardness ratios, optical/near infra-red (NIR)/radio cross-matches, proximity to galaxies, and Galactic positions to improve the classification accuracy. We begin by describing each of these features and Table \ref{tab:other_features} is a summary of the contextual features used in this paper. 

\begin{table*}
\caption{List of contextual features used for classification} 
\label{tab:other_features}
\begin{center} 
\begin{tabular}{lp{12cm}}
\hline
Feature & Description \\
\hline 
\hline
HR1 & $(R_2-R_1)/(R_2+R_1)$ where $R_1$ and $R_2$ are the count rates in the $0.2-0.5$ keV and $0.5-1.0$ keV bands, respectively\\
HR2 & $(R_3-R_2)/(R_3+R_2)$ where $R_2$ and $R_3$ are the count rates in the $0.5-1.0$ keV and $1.0-2.0$ keV bands, respectively\\
HR3 & $(R_4-R_3)/(R_4+R_3)$ where $R_3$ and $R_4$ are the count rates in the $1.0-2.0$ keV and $2.0-4.5$ keV bands, respectively\\
HR4 & $(R_5-R_4)/(R_5+R_4)$ where $R_4$ and $R_5$ are the count rates in the $2.0-4.5$ keV and $4.5-12.0$ keV bands, respectively\\
Bmag & B-band magnitude \\
Vmag & V-band magnitude \\
Rmag & R-band magnitude \\
B-V & B-band magnitude minus V-band magnitude \\ 
Hmag & H-band magnitude \\
Jmag & J-band magnitude \\
Kmag & K-band magnitude \\
J-H & J-band magnitude minus H-band magnitude \\
J-K & J-band magnitude minus K-band magnitude \\
Optical Bayes & Bayes factor for optical cross-match\\ 
Radio & Radio flux at either 1.4\,GHz or 843\,MHz, depending on whether a NVSS or a SUMSS match was found, respectively\\
Radio Bayes & Bayes factor for radio cross-match \\ 
isGalaxyAssociation & Whether there is a galaxy association (Yes or No) \\ 
Luminosity & If there is a galaxy association, then calculate the X-ray luminosity of the source using the galaxy's distance \\
r\_ratio & If there is a galaxy association, distance to galaxy center divided by radius of the galaxy \\ 
galAngSep & Angular separation between the centroid of a galaxy and the position of the source \\
Gal\_lat & Galactic latitude \\
Gal\_lon & Galactic longitude \\
\hline 
\end{tabular}
\end{center} 

\end{table*} 

\subsection{Description of features} 
 
\subsubsection{Hardness ratios} 

Hardness ratio is a crude proxy for the shape of the X-ray spectrum and it has been used with moderate success to classify X-ray sources \citep{kahabka1999}. The \textit{XMM-Newton} EPIC cameras cover the energy band from 0.2~keV to 12.0~keV. The photons gathered are separated into five bands by the 2XMM pipeline, from which four hardness ratios are calculated as follows:  

\begin{equation} 
HR_n = (R_{n+1}-R_n)/(R_{n+1}+R_n) 
\end{equation} 

where $R_n$ is the count rate in the $n$th energy band (see Table \ref{tab:other_features} for the energy range covered by each band). If both bands have count rates within $3\sigma$ of zero, the resulting hardness ratio can be unpredictable as one is essentially dividing one very small number by another very small number. For these cases, we set the hardness ratio to -10.0 as a flag. 

\subsubsection{Optical/NIR cross-matches}

For optical and NIR cross-matching, we used the Naval Observatory Merged Astrometric Dataset  \citep[NOMAD; ][]{zacharias2004}. NOMAD is a conglomeration of various optical photometry and astrometry catalogs and the near-infrared 2MASS catalog. 

To estimate the probability of a chance cross-match, we used the Bayesian method from \citet{budavari2008} where we compared the hypothesis that the cross match is genuine to the alternate hypothesis that the source and the optical counterpart are two unrelated sources. The ratio of the likelihood of these two hypotheses is known as the Bayes factor, $\mathscr{B}$, given by the formula: 

\begin{equation} 
\label{eq:bayes} 
\mathscr{B} = \frac{2}{\psi_1^2+\psi_2^2} exp\left[-\frac{\phi^2}{2\left(\psi_1^2+\psi_2^2\right)}\right],
\end{equation} 
where $\psi_1$ and $\psi_2$ are the resolution of the two catalogs in arcsec and $\phi$ is the angular separation between the two sources. A high Bayes factor favors the hypothesis that the cross-match is genuine. This calculation does not take into account the sky density of the optical sources. For our feature set, we included the B, V, J, H, K band magnitudes if a cross match was found, and the corresponding Bayes factor. If no cross-match was found, the magnitude was set to 100 as a null flag. 

\subsubsection{Radio cross-matches} 

We cross-matched the 2XMMi sources with three radio catalogs --- the NRAO VLA Sky Survey \citep[NVSS; ][]{condon1998}, the Sydney University Molonglo Sky Survey \citep[SUMSS; ][]{mauch2003}, and the Second Epoch Molonglo Galactic Plane Survey \citep[MGPS-2; ][]{murphy2007}. Together, these catalogs provide all-sky coverage of the radio sky. NVSS was a 1.4\,GHz radio survey with the Very Large Array covering the entire sky north of declination -40 degrees. SUMSS was the counterpart survey with the Molonglo telescope of the southern sky (south of declination -30 degrees) at 843\,MHz; MGPS-2 was the Galactic plane radio survey at the same frequency. The positional accuracy of NVSS is $<1\arcsec$  for sources stronger than 15~mJy, and $7\arcsec$ in the survey limit. For SUMSS and MGPS-2, the position accuracy is poorer but typically better than $5$\arcsec. Since the angular resolution of \textit{XMM-Newton} EPIC is better than those of NVSS or SUMSS, for our cross-matching we used a $3\sigma$ search radius based on the radio catalogs. We also included the Bayes factor (Equation \ref{eq:bayes}) to estimate the likelihood of a cross-match. The relatively low sky density of radio sources means that a spurious match is unlikely. 

\subsubsection{Associations with galaxies} 

X-ray sources that correspond to the nuclei of galaxies are likely to be AGN, whilst non-nuclear extragalactic X-ray sources with luminosities of more than $10^{39}$ ergs$^{-1}$ are potential ULX candidates, but can also be foreground stars, XRBs, CVs, or background AGN. We cross-matched the 2XMMi sources with the Third Reference Catalog (RC3) of galaxies \citep{vaucouleurs1991} to find possible galaxy associations. RC3 contains more than 23,000 galaxies, including almost all galaxies with apparent diameters greater than $1\arcmin$. RC3 contains information on the galaxy center position, the major and minor diameters of the $D_{25}$ isophote (roughly the domain of the galaxy) as well as the position angle. We determined $\alpha$, the ratio between the angular separation between the source and the galaxy center and the elliptical radius $R_{25}$. If $\alpha < 1.5$, then we considered the source to be associated with the galaxy. For sources associated with a galaxy, we included $\alpha$ and the angular separation in the feature set. 

\subsubsection{Galactic coordinates} 

\begin{figure} 
\begin{center} 
\includegraphics[width=0.5\textwidth]{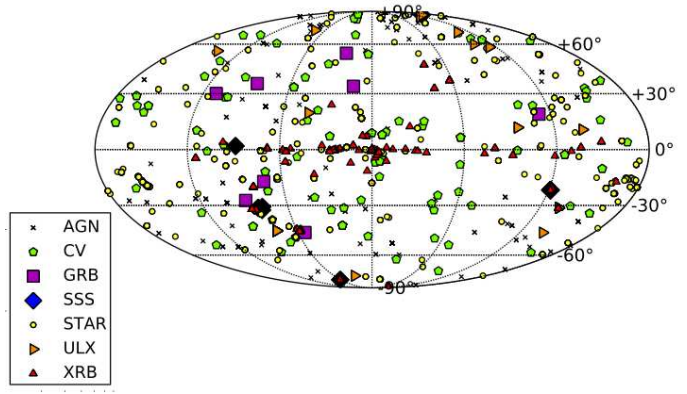} 
\caption{Distribution of sources in our training set in Galactic coordinates.} 
\label{fig:gal_coord}
\end{center} 
\end{figure}

The last set of features we included is the Galactic position of each source. From Figure \ref{fig:gal_coord}, it can be seen that XRBs are more likely to cluster along the Galactic plane, whilst all other source types are distributed isotropically in Galactic coordinates. This motivates the inclusion of Galactic coordinates in the feature set as a way to identify XRBs. 

\subsection{Accuracy of training set} 

\begin{figure} 
\begin{center} 
\includegraphics[width=0.5\textwidth]{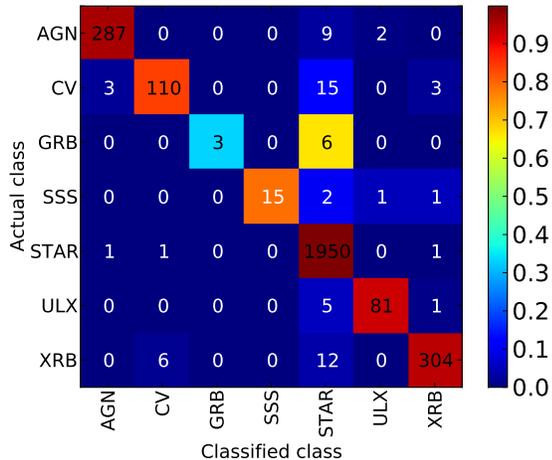} 
\caption{Confusion matrix from performing 10-fold cross-validation on the training set using the RF classifier with time-series and contextual features. The color bar represents the true positive rate. The overall accuracy is 97\%.} 
\label{fig:conf_training_all}
\end{center} 
\end{figure}

\begin{figure} 
\begin{center} 
\includegraphics[height=0.4\textwidth]{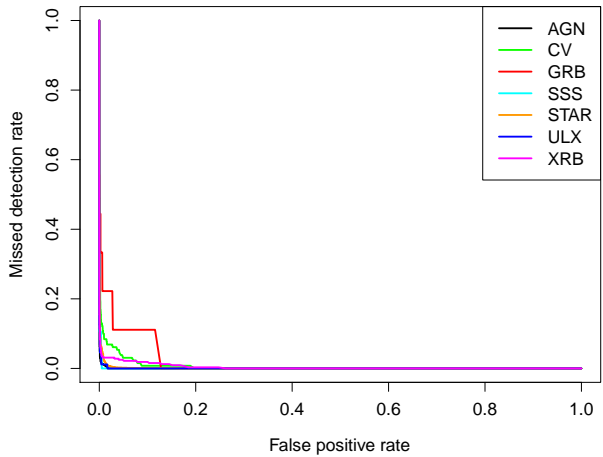} 
\caption{ROC plot from performing 10-fold cross-validation on the training set using the RF classifier with time-series and contextual features. Performance across all classes show marked improvement from classification using only time-series features.} 
\label{fig:roc_training_all}
\end{center} 
\end{figure}

As in Section \ref{sec:acc_timeseries}, we used 10-fold cross-validation to evaluate the performance of this feature set. Using both the time-series and contextual feature sets, the overall accuracy improved significantly from 77\% to 97\% with the additional features. Figures \ref{fig:conf_training_all} and \ref{fig:roc_training_all} show the confusion matrix and the ROC plot, respectively. Performance improved across all classes relative to Figures \ref{fig:conf_training} and \ref{fig:roc_training}. However, we misclassified most GRBs as stars (Figure \ref{fig:conf_training_all}). From the ROC plot, it can be seen that we can achieve 100\% accuracy for GRBs if we are willing to accept a 5\% false positive rate. Even though this does not seem high, it would mean an extra ${\sim}140$ sources misclassified as GRBs in exchange for accurately classifying 9 GRBs. We will discuss this issue with minority classes in more detail in Section \ref{s_dr3}. 

\begin{figure} 
\begin{center} 
\includegraphics[height=0.6\textwidth]{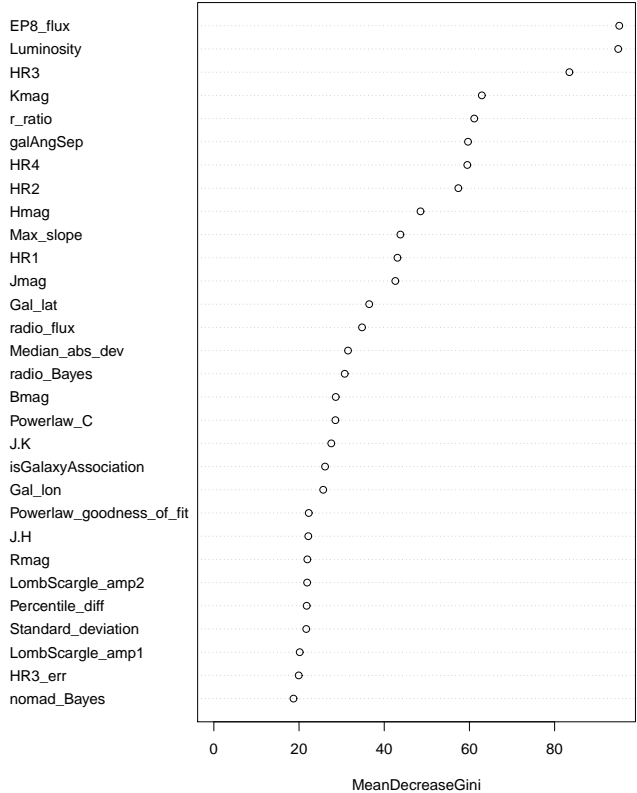} 
\caption{Relative importance of the time series and contextual features. The features are described in detail in Tables \ref{tab:ts_features} and \ref{tab:other_features}. } 
\label{fig:varImp_all}
\end{center} 
\end{figure}

Figure \ref{fig:varImp_all} shows the relative feature importance of the top 30 features, determined using the mean decrease in Gini impurity. X-ray flux and X-ray luminosity (for sources with a galaxy association) are the most informative features, followed by HR3. Overall, hardness ratios appear to be highly informative, with all four hardness ratios placed in the top 10 of most informative features. On the other hand, time-series features do not rank highly on the list. 

\section{Catalog of probabilistically classified XMM variable sources} \label{s_unknown}

\subsection{Results} 

Using the entire training set, we constructed a RF classification model using the method described in Section \ref{s_contextual}. Then we applied this classification model to the set of unknown 2XMMi variable sources. For sources where there are more than one detection, we classified each detection separately and combined the results by averaging the output class membership probabilities. Table \ref{tab:results_overview} shows the number of unknown sources classified as one of seven classes. The majority of the unknown sources are classified as stars. 

We also compiled a downloadable table of the class membership probabilities. Table \ref{tab:xmm_tab} shows a portion of that table. 
 
\begin{table}
\caption{Unknown variable source classification} 
\label{tab:results_overview}
\begin{center} 
\begin{tabular}{lr} 
Class & Number \\
\hline 
\hline 

AGN  & 25 \\
CV   & 3 \\
GRB  & 0 \\
SSS  & 0 \\
STAR & 362 \\
ULX  & 7 \\
XRB  & 14 \\
\hline 
Total & 411\\
\hline 
\end{tabular}
\end{center} 
\end{table} 

\begin{table*}
\begin{center}
\begin{threeparttable}
\caption{2XMM variable sources classification} 
\label{tab:xmm_tab}  
\begin{tabular}{ccccccccccc} 
2XMM name & P$_{\textrm{AGN}}$ & P$_{\textrm{CV}}$ & P$_{\textrm{GRB}}$ & P$_{\textrm{SSS}}$ & P$_{\textrm{Star}}$ & P$_{\textrm{ULX}}$ & P$_{\textrm{XRB}}$ & P$_{\textrm{max}}$ & Output class & Margin \\
\hline
\hline
J000028.5$+$622220&0.000&0.004&0.000&0.000&0.992&0.000&0.004&0.992&Star&0.984 \\
J000354.2$-$255841&0.078&0.014&0.048&0.001&0.810&0.005&0.044&0.810&Star&0.620\\
J000511.6$+$634018&0.133&0.027&0.009&0.001&0.786&0.001&0.043&0.786&Star&0.572\\
J002111.3$-$084140&0.005&0.018&0.000&0.000&0.970&0.000&0.007&0.970&Star&0.940\\
J002133.3$-$150751&0.502&0.025&0.060&0.007&0.346&0.010&0.050&0.502&AGN&0.004\\
J002523.8$+$640932&0.130&0.166&0.030&0.047&0.438&0.047&0.142&0.438&Star&-0.124\\
J002525.5$+$640821&0.026&0.098&0.093&0.001&0.681&0.063&0.038&0.681&Star&0.362\\
J004134.2$+$851230&0.122&0.053&0.007&0.002&0.788&0.003&0.025&0.788&Star&0.576\\
J004146.7$+$402421&0.100&0.049&0.011&0.000&0.828&0.001&0.011&0.828&Star&0.656\\
J004213.0$+$411835&0.040&0.002&0.000&0.022&0.019&0.027&0.890&0.890&XRB&0.779\\
J004218.3$+$411223&0.052&0.002&0.001&0.131&0.081&0.046&0.687&0.687&XRB&0.375\\
J004221.4$+$411600&0.032&0.001&0.000&0.041&0.016&0.009&0.901&0.901&XRB&0.801\\
J004233.8$+$411619&0.166&0.002&0.000&0.158&0.150&0.132&0.391&0.391&XRB&-0.218\\
J004508.5$+$421546&0.284&0.024&0.024&0.001&0.637&0.004&0.026&0.637&Star&0.274\\
J005759.9$-$272126&0.033&0.012&0.000&0.000&0.949&0.000&0.006&0.949&Star&0.898\\
J011543.9$+$330845&0.020&0.055&0.003&0.001&0.912&0.002&0.007&0.912&Star&0.824\\
J011615.3$-$732655&0.267&0.024&0.016&0.002&0.356&0.025&0.311&0.356&Star&-0.288\\
J011722.4$-$732418&0.214&0.035&0.042&0.000&0.661&0.002&0.046&0.661&Star&0.322\\
J012539.7$+$315511&0.293&0.041&0.040&0.009&0.573&0.004&0.040&0.573&Star&0.146\\
J013324.3$+$304402&0.005&0.001&0.000&0.003&0.005&0.015&0.971&0.971&XRB&0.941\\

\hline 
\end{tabular} 
\begin{tablenotes}[para] 
{\bf Notes.}\\
This is the first 20 lines of the table. The complete table is available electronically. 
Column 1: 2XMM name. Columns 2-8: Probability given by our Random Forest classifier that the source belongs to the class AGN, CV, GRB, SSS, Star, ULX and XRB respectively. Column 9: Probability given to the output class. Column 10: Class given to the source by our classifier. Column 11. Classification margin calculated as probability of output class minus probability of not belonging to output class. Margin is another way to state P$_{max}$ with margin = 2P$_{max}$ - 1. Probabilities may not add up to 1 due to rounding errors. 
\end{tablenotes} 
\end{threeparttable}
\end{center}
\end{table*}

\subsection{Evaluation of results} 
\subsubsection{Comparison with recent classifications in literature} 
\label{sec:recent_class} 

Following the initial source classification (Farrell et. al., in prep), a number of sources in the unknown sample have since been classified in the literature. We assessed the accuracy of the classifier by comparing the literature classification to the output of our RF classifier for ${\sim}12\%$ of the the unknown sources. Confirming the classification for 411 X-ray sources is beyond the scope of this paper. We found recently confirmed or tentative classifications for 19 sources and they are listed in Table \ref{tab:recent_classification}. The classifications from our RF classifier agree with the literature classifications in 13 out of 19 cases if we include the two sources that have multiple possible classifications. The misclassifications are due to the source belonging to a novel source type, insufficient information, poor data quality, or problems with the classification in the literature. Of the six misclassifications, three sources have been classified as ULXs by our RF classifier whilst \citep{kamizasa2012} regarded them as candidate AGN with immediate-mass black holes based on the presence of X-ray variability. The criteria used by \citep{kamizasa2012} do not preclude ULXs since they only filtered out sources in known star forming regions and included sources with object type Galaxy shown in the NED databases. All three of the sources classified as ULXs are close to a galaxy in RC3 and have X-ray luminosity of between $10^{39}$ and $10^{40}$ ergs/s. Here we briefly discuss three of the other misclassifications. \\

{\bf 2XMM J034645.4$+$680947:}
This source is classified as an XRB by our classifier but \citet{mak2011} classified it as a SSS. However, there are a few problems with the literature classification. \citet{mak2011} only used two hardness ratios in the classification. This is coarser than what we have used, which would have resulted in the loss of information. There are four observations of this source and the hardness ratio only satisfied the criteria for SSS \citep[as defined by][]{mak2011} in the two fainter observations. The lack of X-ray flux in the $2-7$\,keV band could be a selection effect since the hard emission tends to be undetectable in fainter sources. Furthermore, the hardness ratios derived from the 2004 August and 2004 February observations do not classify this source as a SSS. We fitted the 2004 August EPIC spectra that were automatically extracted by the XMM pipeline with a Raymond-Smith model \citep{raymond1977}, typical for a SSS. The best fit parameters are: $N_H = (0.03 \pm 0.03) \times 10^{22}$ cm$^{-2}$, kT = (0.79 $\pm 0.05) $ keV and $\chi^2$ / dof =  175.03/183. This is a satisfactory fit, however the temperature is an order of magnitude higher than typical for a SSS  \citep[SSSs peak in the range $20-100$\,eV;][]{kahabka2006}. From the above arguments, we are skeptical that 2XMM J034645.4$+$680947 is a SSS. Our RF model classified this source as an XRB, SSS, star and ULX with probabilities $0.349$, $0.227$, $0.222$ and $0.16$ respectively. This suggests that either we lack sufficient information to classify this source and/or that this source is highly unusual. \\

{\bf 2XMM J060636.4$-$694937:} 
This source is classified as an AGN by our classifier but has been confirmed as a classical nova in the Large Magellanic Cloud \citep{read2009}. The observation in our sample occurred during the nova outburst phase. Although novae are a subset of CVs we do not have many examples of novae in outburst in the training set, hence to our classifier this is a novel source type. This highlights one of the limitations of supervised classification in that the classifier is incapable of recognizing novel classes. \\

{\bf 2XMM J174016.0$-$290337:}  
This source is classified as an XRB by our classifier. Using only X-ray timing and X-ray spectral data, \citet{farrell2010} identified this source as likely to be a symbiotic XRB, a new and rare sub-class of XRBs composed of a late-type giant accreting matter onto a compact object such as a neutron star. However, with more optical spectral data, \citet{masetti2012} later identified it as an mCV. There is an optical counterpart in the \textit{XMM-Newton} error circle with a spectrum that contains strong Balmer, He I, He II, and Bowen blend emissions, typical of magnetic CVs. Similar to the conclusion made by \citet{farrell2010}, our classifier favors the interpretation of this source as a XRB, giving it a probability of 0.46, but also gives the probability of this being a CV as 0.29. This demonstrates that our classifier is capable of making a conclusion along the same line as an expert in the field using the same information. It is worth noting that this is an unusual source and its X-ray properties do not fit with the interpretation of it being an mCV. \\

\begin{table*}
\begin{center}
\begin{threeparttable} 
\caption{Comparison between RF classification and recent literature classifications} 
\label{tab:recent_classification}
\begin{tabular}{lccccl} 
\hline 
2XMM name & Our classification & Margin\tnote{1} & Literature classification & Confidence \tnote{2} & References \tnote{3} \\
\hline 
\hline

J002133.3$-$150751 & AGN & 0.00 & AGN & T & KTH2012\\
J013612.5$+$154957 & AGN & -0.42 & AGN & T & KTH2012\\
J022428.8$-$041414 & AGN & 0.12 & AGN & C & SPS2010 \\
J023213.4$-$072945 & Star & -0.02 & AGN & T & KTH2012\\
J034645.4$+$680947 & XRB & -0.32 & SSS\tnote{4} & T & MPK2011\\
J060636.4$-$694937 & AGN & -0.06 & CV  & C & RSF2009\\
J064217.8$+$821719 & AGN & -0.09 & AGN & T & AG2007\\
J120143.6$-$184857 & ULX & -0.27 & AGN & T & KTH2012\\
J123103.2$+$110648 & AGN & -0.06 & AGN & T & KTH2012\\
J123316.6$+$000512 & ULX & -0.40 & AGN & T & KTH2012\\
J130543.9$+$181355 & AGN & 0.38 & AGN & T & KTH2012\\
J134736.4$+$173404 & AGN & -0.17 & AGN & T & KTH2012\\
J174016.0$-$290337 & XRB & -0.22 & XRB / CV  & U & MNP2012, FGW2010 \\
J174445.4$-$295046 & XRB & 0.14 & XRB / CV  & U & HTY2009\\
J185330.6$-$012815 & CV & -0.24 & CV   & C & HSC2012 \\
J191043.4$+$091629 & XRB & 0.24 & XRB  & U & PBF2011 \\
J213152.8$-$425130 & ULX & -0.07 & AGN & T & KTH2012\\
J233430.3$+$392101 & AGN & 0.56 & AGN & T & KTH2012\\
J235509.6$+$060041 & AGN & 0.66 & AGN & T & KTH2012\\
\hline 
\end{tabular}
\begin{tablenotes}[para] 
{\bf Notes.} \\
 \item[1] {Margin refers to output probability for the given class minus the probability that it is not the given class.}; 
 \item[2] {Confidence of the classification given in the literature. C - confirmed, T - tentative, U - uncertain, multiple source types are consistent with the available data}; 
 \item[3] {KTH2012 - \citet{kamizasa2012}, SPS2010 - \citet{stalin2010}, MPK2011 - \citet{mak2011}, RSF2009 - \citet{read2009}, AG2007 - \citet{atlee2007}, MNP2012 - \citet{masetti2012}, FGW2010 - \citet{farrell2010}, HTY2009 - \citet{heinke2009}, HSC2012 - \citet{hui2012}, PBF2011 - \citet{pavan2011}};
 \item[4] {We are skeptical of the classification of this source as SSS. See text for more details.} 
\end{tablenotes}
\end{threeparttable} 
\end{center}
\end{table*} 

\subsubsection{Classification of known sources in 2XMMi-DR3} \label{s_dr3}

\begin{table*}
\begin{center}
\begin{threeparttable} 
\caption{DR3 target source classifications} 
\label{tab:dr3_classification}
\begin{tabular}{llccccc} 
\hline 
2XMM name & Other name & Literature classification & Our classification & Margin \tnote{1} \\
\hline 
\hline
J002257.7$+$614107&IGR00234$+$6141  &CV&CV                     & 0.72  \\
J005519.7$+$461257&XSS0056$+$4548   &CV&CV                     & 0.51  \\
J023155.2$-$711806 &GRB080411              &GRB&GRB              &  0.50  \\
J050106.5$+$451634 &SGR 0501$+$4516 &Magnetar&XRB     & 0.04  \\
J051045.5$+$162958 &IRAS05078$+$1626&AGN&AGN            & -0.38  \\
J052031.8$+$061611&RX J0520.5$+$0616&Star&Star                & 0.72  \\
J053450.5$-$580141&TW Pic                      &CV&CV                       & 0.77  \\
J061322.3$+$474425&SS Aur                     &CV&CV                       & 0.00  \\
J084047.8$-$450329&IGR J08408$-$4503 &XRB&Star               & 0.83  \\
J114720.0$-$601427 &GRB080723B           &GRB&GRB               & -0.16  \\
J123907.8$-$453344&HD 109962                &Star&Star                 & 0.75  \\
J132344.6$-$414429&V803 Cen                 &Star&Star                   &  -0.14  \\
J141922.3$-$263841&ESO511$-$G030       &AGN&AGN             & 0.86  \\
J143104.7$+$281713&MRK 684                  &AGN&AGN               & 0.58  \\
J144244.4$-$003956&RXJ 1442$-$0039     &Star&Star                & 0.82  \\
J161534.5$-$224241&VV Sco                       &Star&Star                 & 0.91  \\
J164547.6$-$453642&GX 340$+$0              &XRB&XRB               & 0.88  \\
J172840.4$-$465349&GJ 674                      &Star&Star                   & 0.40  \\
J173911.5$-$302036&XTE J17391$-$3021 &XRB&XRB              & -0.02  \\
J174009.1$-$284725 &AX J1740.1$-$2847 &XRB&CV              & -0.15 \\
J180339.6$+$401220&RXJ 1803$+$4012  &CV&CV                   & 0.56  \\
J181613.3$+$495203&AM Her                    &CV&CV                     & 0.23  \\
J183219.3$-$084030&AX J1832.3$-$0840 &CV&XRB               & 0.02  \\
J184508.4$-$635742&SCR J1845$-$6357 &Star&Star              & 0.43  \\
J194301.7$+$321912 &V2491 Cygni            &CV&XRB                & -0.45  \\
J195436.5$+$322155&EY Cyg                    &CV&CV                & -0.17  \\
J221402.5$+$124211 &RU Peg                    &CV&XRB           & -0.14  \\

\hline 
\end{tabular}
\begin{tablenotes}[para] 
{\bf Notes.} \\
 \item[1] {Margin refers to output probability for the given class minus the probability that it is not the given class.}
\end{tablenotes}
\end{threeparttable} 
\end{center}
\end{table*} 

Another method we used to evaluate the performance of our classifier is to use the classification model to classify 27 known variable sources in 2XMMi-DR3 that were not in DR2. The 2XMMi-DR3 catalog is an incremental update to the 2XMMi-DR2 catalog and consists of all of 2XMMi-DR2 plus observations made between August 2008 and October 2009. The 27 sources we have chosen are the targets of observations with known classifications, but which are not in our training set. We classified 22 out of 27 sources correctly, which gives an accuracy of ${\sim}81\%$ (Table \ref{tab:dr3_classification}). This is lower than the accuracy from 10-fold cross-validation of ${\sim}97\%$. However, this is not unexpected since the composition of source types in this DR3 subset is vastly different to the training set. For instance, 37\% of sources in the DR3 subset are CVs but in our training set, only 4.6\% are CVs. In the 10-fold cross-validation, 69\% of sources are stars for which the classification accuracy is 99.8\%. We were able to classify all seven stars in the DR3 subset correctly. 

Of the seven DR3 sources that we mis-classified, two are unusual sources. 2XMMi J050106.5$+$451634 is a known magnetar \citep{rea2009}, a source type that is not in our training set. Although there were magnetars in the variable samples, we excluded them from the training set since there were only very few samples. The other source, 2XMMi J084047.8$-$450329, is a recurrent supergiant fast X-ray transient \citep[SFXT;][]{leyder2007}. SFXTs have only been identified recently as a new class of XRBs and are believed to consist of a wind-accreting compact object and an OB super-giant donor star. In both cases, our classifier was not able to devise a correct classification because the correct class is not one that the classifier has knowledge of. 

There are two GRBs in our DR3 subset and we correctly classified both of them, despite GRB being a minority class. We repeated the experiment and trained a RF classifier without resampling the data set and found that it was only able to identify one of the two GRBs. This demonstrates that resampling is important for achieving good performance on minority classes. 

\section{Anomalous sources} \label{s_interesting}

\begin{figure} 
\begin{center} 
\includegraphics[width=0.45\textwidth]{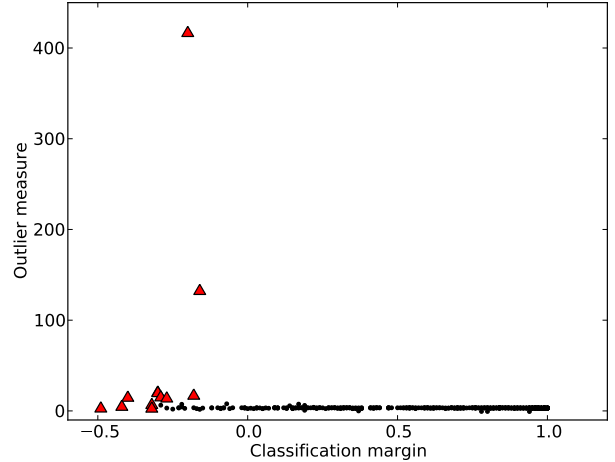} 
\caption{Outlier measure versus classification margin for the unknown variable 2XMM sources. The red triangles are sources that have an outlier measure greater than 10 or classification margins less than $-0.3$. Table \ref{tab:outlier} contains more detail on the potential outlier sources.} 
\label{fig:outlier}
\end{center} 
\end{figure} 

In the previous section, we used mislabelled instances to highlight one of the issues with supervised classification - namely that it cannot label novel source types. Uncovering novel but rare source types is a stated goal of many large surveys. In machine learning, this task is known as anomaly detection \citep{chandola2009}. Anomalies are cases whose proximity to other cases of the same type is small. In the \texttt{R} package \texttt{randomForest}, there is a function to calculate an outlier measure based on the proximity matrix. The proximity matrix $Prox$ is an $S$ by $S$ matrix (assuming the training set has $S$ cases) where $Prox(i,j)$ is incremented by one if case $i$ and case $j$ both end up in the same terminal node of a tree. $Prox$ is normalized by dividing by the number of trees in the forest. The outlier measure calculates the proximity of the case $i$ to other cases of the same class, using the equation: 

\begin{equation}
\label{eqn:outlier} 
O(i) = \frac{1}{\rm{MAD}} \left( \frac{n}{ \sum_{class(j) = k}Prox^2(i,j)} - \mathscr{M} \right),
\end{equation} 

where $k$ is the class of case $i$, and $n$ is the number of instances of the class $k$. $O(i)$ is normalised by subtracting $\mathscr{M}$, the median of the unnormalised outlier measures, and dividing by the median absolute deviation ($\rm{MAD}$). Higher outlier measures mean the source is more anomalous whilst a low outlier measure means the source is similar to other sources of the same class. 

\begin{table*}
\begin{center}
\begin{threeparttable} 
\caption{Anomalous unknown sources} 
\label{tab:outlier}
\begin{tabular}{llccrcrl} 
\hline 
Number & 2XMM name & Our classification & Margin & Outlier & Sum flag & S/R & Notes \\
\hline 
\hline

1&J013612.5$+$154957&AGN&-0.42&4.4&1&23&AGN candidate \citep{kamizasa2012}\\
2&J034645.4$+$680947&XRB&-0.32&6.4&0&31&Anomalous source, see Section 6.1\\
3&J045445.5$-$180641&ULX&-0.30&19.5&3&7& Low S/N\\
4&J120143.6$-$184857&ULX&-0.27&13.6&0&22&AGN candidate \citep{kamizasa2012}\\
5&J122543.2$+$333253&ULX&-0.30&20.1&4&14&Next to a bright source\\
6&J122549.0$+$333202&ULX&-0.18&16.7&3&23&Next to a bright source\\
7&J123316.6$+$000512&ULX&-0.40&14.4&0&18&AGN candidate \citep{kamizasa2012}\\
8&J155013.0$-$034749&CV&-0.16&132.1&0&20&Large flare\\
9&J161741.9$-$833751&AGN&-0.32&2.4&1&51&Stochastic X-ray variability\\
10&J180658.7$-$500250&AGN&-0.49&2.6&3&211&Anomalous source, see Section 7.1\\
11&J181330.6$-$333627&CV&-0.20&416.4&3&49&V2694 pulsating star\\
12&J231818.7$-$422237&ULX&-0.29&14.86&1&32&Next to a bright source\\

\hline 
\end{tabular}
\begin{tablenotes}[para] 
{\bf Notes.} \\
Column definitions from the left: Source number; 2XMM name; classification given to the source by our RF classifier; classification margin; outlier measure (Equation \ref{eqn:outlier}); sum flag which is a measure of data quality from the XMM data processing pipeline (scale of 0 to 4; 0 means good, 4 means source is possibly spurious); signal-to-noise ratio; notes 
\end{tablenotes}
\end{threeparttable} 
\end{center}
\end{table*} 

Figure \ref{fig:outlier} is a plot of the classification margin against the outlier measure for the 408 unknown sources in our test set, excluding the three sources with recent spectroscopic identifications listed in Table \ref{tab:recent_classification}. The sources marked as red triangles either have outlier measures greater than 10 and/or classification margins of less than $-0.3$, making them likely to be true outliers. The cut-off is arbitrary and we use it to select a manageable number of potential outlier sources to verify this technique. 12 sources satisfy these criteria and are listed in Table \ref{tab:outlier}. We now discuss possible reasons why these sources have been deemed anomalous. 

A common reason for sources to be classified as anomalous is bad data quality. For sources 3, 5, 6, and 12, the source of interest is either very close to an extremely bright source, or within the confines of a diffuse source. These situations can lead to the contamination of the X-ray spectrum (thereby giving unreliable fluxes and HRs) and/or time series. Sources with low signal-to-noise ratios (S/N) can also be erroneously classified as anomalous. Source 3 has S/N of less than 10 and we cannot trust the classifier's determination that it is anomalous. Low S/N means that the hardness ratios will have large error bars and that errors on features will not be used properly in the classification process. Incorporating error bars into the classification algorithm is an area to be addressed in future work. 

One of the anomalous sources on our list is source 2, 2XMM J034645.4$+$680947, which we have already discussed in Section \ref{sec:recent_class}. We will now briefly consider the nature of source 12. \\

{\bf 2XMM J180658.7-500250 (source 12):}
\label{sec:weird} 

\begin{figure} 
\begin{center} 
\includegraphics[width=0.5\textwidth]{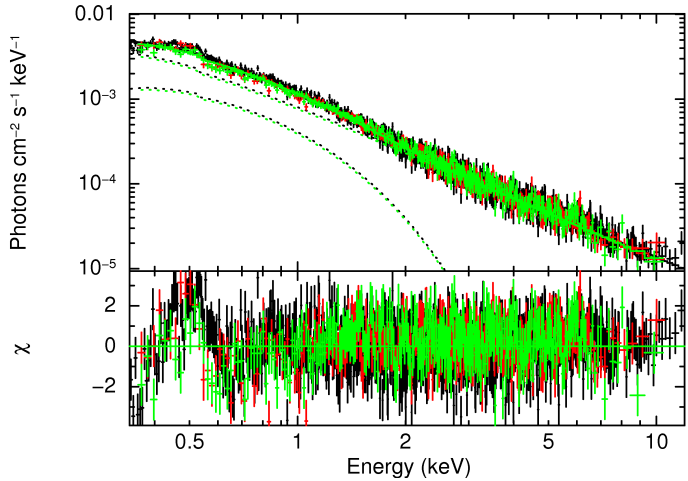} 
\caption{Fit of the X-ray spectrum of 2XMM J180658.7$-$500250 using an absorbed power-law + disk black-body model. The top panel shows the spectrum (solid lines) and the fit (dotted lines) -- red is MOS1, green is MOS2 and black is pn; curved line is the disk black-body model, straight line is the power law model. The bottom panel shows the residuals from the fit.} 
\label{fig:weird_spectrum}
\end{center} 
\end{figure} 

\begin{figure} 
\begin{center} 
\includegraphics[width=0.45\textwidth]{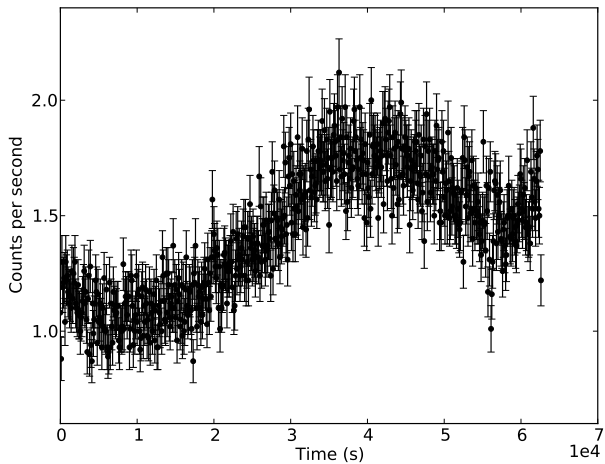} 
\caption{pn light curve of 2XMM J180658.7$-$500250} 
\label{fig:weird_lc}
\end{center}
\end{figure} 

This is the most unusual source in our list. It has a counterpart in 2MASS with J = 14.144 mag, H = 13.446 mag, and K = 12.715 mag, but no other optical or radio counterparts were found in the literature\footnote{The absence of an optical counterpart is not surprising considering that 2XMM J180658.7-500250 is very close in terms of angular distance to the nearby bright star Theta Ara, making it very difficult to obtain accurate photometry. }. There is also a mid-IR match from the Wide-field Infrared Survey Explorer (WISE) survey with magnitudes at 12 and 22 $\mu$m of 7.356 and 5.165, respectively. The mid-IR WISE colors are consistent with a spiral galaxy \citep{wright2010}, indicating that this object may be extragalactic. From the WISE image this source looks unresolved, which gives it an upper limit angular size of $10\arcsec$, and if we assume the size of a spiral galaxy to be at least 5\,kpc, then we can constrain the distance to this source to be at least 100\,Mpc. 

We attempted to fit the X-ray spectra with different simple one-component models (power law, black-body, disk black-body, thermal plasma, and bremsstrahlung models). The absorbed power-law model produced the best fit with $\chi^2$/dof = 1888.43/1417. However, as this is not a statistically acceptable fit we tried more complex two component models (power law + disk black body, power law + black body, mekal + mekal, bremsstrahlung + blackbody). The two component model that provided the best physical fit to the data is the absorbed power law + disk black body model (Figure \ref{fig:outlier}). The best fit parameters are: nH = $2.9\times10^{20}$ atoms cm$^{-2}$, $\Gamma = 1.8$, T = 0.36 keV. However, it is still not a good statistical fit, with $\chi^2$/dof = 1702.88/1415 and significant residuals around 0.5 keV (see Figure \ref{fig:weird_spectrum}).

Such low-energy features are reminiscent of those seen in the spectra of the well known ULX Holmberg II X-1. Using high-quality spectra obtained with the XMM-Newton EPIC and Reflection Grating Spectrometer (RGS) instruments, \citet{goad2006} detected evidence for a complex of emission lines in the spectra of Holmberg II X-1 with energies between 563 -- 577 eV, possibly associated with the O VII triplet. We therefore added a Gaussian component to our best fit absorbed power law plus disk black body model, finding that a broad emission line (E$_{Gauss}$ = 0.50 $\pm$ 0.01 keV, $\sigma$ = 0.05 $\pm$ 0.01 keV, equivalent width = 0.044 keV, and unabsorbed bolometric flux = (1.5$^{+0.7}_{-0.4}$) $\times$ 10$^{-13}$ erg cm$^{-2}$ s${-1}$) improved the fit significantly to a statistically acceptable $\chi^2$/dof = 1530.77/1412. In the case of Holmberg II X-1, \citet{goad2006} found evidence for the presence of two narrow O VII lines: the forbidden line at 563 eV and the resonance line at 577 eV. However, replacing our broad line with two narrow lines worsened our fit. \citet{goad2006} also considered the possibility that the emission features they detected were due to the presence of an optically thin thermal plasma. We thus tried refitting our spectra of 2XMM J180658.7$-$500250 with the MEKAL thermal plasma model replacing the disk black body and Gaussian emission features. However, this did not improve the fit ($\chi^2$/dof = 1678.46/1414).

The possible coincidence with a spiral galaxy at a redshift between z $\sim$ 0.01 -- 0.05 combined with the spectrum reminiscent of Holmberg II X-1 raises the possibility that 2XMM J180658.7$-$500250 may also be a ULX. If this is the case, the absorbed 0.2-10 keV flux of (6.05$^{+0.05}_{-0.06}$) $\times$ 10$^{-14}$ erg cm$^{-2}$ s$^{-1}$ using the best fit power law plus disk black body plus Gaussian model implies a luminosity between $\sim$10$^{42}$ - 10$^{43}$ erg s$^{-1}$, which would make 2XMM J180658.7$-$500250 even more luminous than the brightest ULX (and strongest intermediate mass black hole candidate) ESO 243-49 HLX-1 \citep{farrell2009}. Such luminosities are extremely difficult to explain with a stellar mass black hole, implying a black hole mass of $\gg$ 1000 M$_\odot$. A luminosity this high is much more reminiscent of those observed from AGN, however the disk black body temperature is too high for an accretion disk around a supermassive black hole. However, such a disk black body temperature is typical of ULXs \citep{berghea2008}. 

In addition, the light curve of this source is also interesting and has a tantalising hint of periodicity (Figure \ref{fig:weird_lc}). Periodic variability does not fit with the AGN interpretation and instead favours the compact binary object classification. In addition, the mid-IR color is highly unusual for an AGN, and is instead much more reminiscent of a non-active galaxy. We therefore speculate that 2XMM J180658.7$-$500250 may be a new member of the extremely rare class of hyper-luminous X-ray sources (HLXs), i.e. ULXs with luminosities in excess of 10$^{41}$ erg s$^{-1}$, that potentially represent the best candidates for intermediate mass black holes.  

In summary, 2XMM J180658.7$-$500250 appears to be unusual source that our classifier has rightly picked out as anomalous. More work, such as fitting more complex X-ray spectral models as well as multi-wavelength follow-up observations, is needed to verify its nature.

\section{Conclusions}\label{s_conclusion} 

In this paper, we have tested the performance of the RF classifier with the 2XMMi-DR2 data set. On a seven class data set with only time series features, we were able to attain a 10-fold validation accuracy of ${\sim}77\%$. Time series features do have some discriminative power, but in the absence of other information, they do not result in a high performing classifier. When we added in contextual features such as hardness ratios,  optical/IR/radio cross-matches, Galactic coordinates and proximity to nearby galaxies, the classification accuracy increased to ${\sim}97\%$. This shows that the RF classifier can be a high performing classifier, but only by combining both time-series and contextual features. The same conclusion was made by \citet{palaversa2013} in their work on the automatic classification of optical stars, in which they found that using both light curve features and colours allowed them to achieve accuracy of 92\%. A potential recommendation from our work is that the classifiers for future synoptic variable surveys will need more than just temporal flux measurements to achieve good performance. 

We demonstrated the scientific potential of an automatic classifier by applying our random forest classifier to 411 unknown variable sources. To test the reliability of such automatic classification, we found recent classifications in the literature for 19 sources and checked the literature's suggested classification against the output from our classifier. Our classification agrees with the literature in 13 out of the 19 sources (accuracy of 68\%). The mislabelled cases are due to a source belonging to a new and unseen class or because the classification made in literature used information (such as optical spectra) that were not available to us. We also used our RF classifier on a known subset of target sources in 2XMM-DR3. We were able to classify 22 out of 27 sources correctly (accuracy of 81\%). The mislabelled sources are again of unknown source types, or are unusual members of one of the known source types.  

In the DR3 verification exercise, we showed that the RF classifier can accurately classify GRBs, a heavily under-represented class. This performance was achieved by oversampling the minority classes.

To find anomalous sources, we used the classification margin and the outlier measure from the RF package. Most of the high potential anomalous sources we found contained data quality issues. One source in our list did look genuinely unusual (2XMM J180658.7$-$500250)  and further work needs to be done to determine its true nature. 

There are two areas for improvement on the algorithm front. First, to the best of our knowledge, current machine learning algorithms (including RF) do not take into account the error bars in the features. In astronomy, accurate measurement errors are readily available and provide valuable information, and should be incorporated into the machine learning algorithm. One simple way to do this is to apply a weighting to reflect the size of the error. This needs to be done in such a way that would propagate the error to the classification accuracy. Second, the RF classifier lacks interpretability. For an individual source, the RF classifier does not allow the user to pinpoint the feature which led to the classification, which is something that a human expert can easily provide. However, RF can provide a measure of feature importance measured using all the samples in the training set. 

Automatic classification will likely play a major role in future synoptic surveys across all wavelengths. In this paper, we have shown that the RF classifier can achieve excellent performance. We envision that a similar model can be built into the pipeline for time-domain surveys on the SKA and the LSST, where the goal will be to produce probabilistic classifications as a value-added component to the catalogs. 

\section*{Acknowledgements} 
This research was conducted by the Australian Research Council Centre of Excellence for All-sky Astrophysics (CAASTRO), through project number CE110001020. K. K. L is supported by a university postgraduate award from the University of Sydney and a scholarship from CAASTRO. S. A. Farrell is the recipient of an Australian Research Council Post Doctoral Fellowship, funded by grant DP110102889. Based on observations from XMM-Newton, an ESA science mission with instruments and contributions directly funded by ESA Member States and NASA. This work made use of the 2XMM Serendipitous Source Catalogue, constructed by the XMM-Newton Survey Science Centre on behalf of ESA. This research has also made use of the NASA/IPAC Extragalactic Database (NED) which is operated by the Jet Propulsion Laboratory, California Institute of Technology, under contract with the National Aeronautics and Space Administration, and the SIMBAD database, operated at CDS, Strasbourg, France. We thank the referee for comments that have helped to improve the paper. 


\bibliographystyle{apj}
\bibliography{journals,xmm_class}




\end{document}